\documentclass[modern]{aastex62}

\newcommand{\simless}{\mathbin{\lower 3pt\hbox
     {$\rlap{\raise 5pt\hbox{$\char'074$}}\mathchar"7218$}}}
\newcommand{\simgreat}{\mathbin{\lower 3pt\hbox
     {$\rlap{\raise 5pt\hbox{$\char'076$}}\mathchar"7218$}}}

\newcommand{\mearth}{\,M$_\oplus$}

\newcommand{\coj}{$^{12}$CO(2-1)}
\newcommand{\kms}{\,km\,s$^{-1}$}

\usepackage{amsmath, amssymb,mathrsfs,mathtools}
\newcommand{\ugc}[1]{}

\usepackage{amsmath }
\usepackage{tablefootnote}

\begin{document}

\title{Modelling the Spatial Distribution and Origin of CO Gas in Debris Disks}

\author{A.S. Hales}
\affiliation{Joint ALMA Observatory, Avenida Alonso de C\'ordova 3107, Vitacura 7630355, Santiago, Chile}
\affiliation{National Radio Astronomy Observatory, 520 Edgemont Road, Charlottesville, VA 22903-2475, United States of America}
\author{Uma Gorti}\affiliation{NASA Ames Research Center, MS 245-3, Moffett Field, CA 94035-1000}
\affiliation{SETI Institute, Mountain View, CA 94043}
\author{John M. Carpenter}
\affiliation{Joint ALMA Observatory, Avenida Alonso de C\'ordova 3107, Vitacura 7630355, Santiago, Chile}
\affiliation{National Radio Astronomy Observatory, 520 Edgemont Road, Charlottesville, VA 22903-2475, United States of America}
\author{Meredith Hughes}\affiliation{Wesleyan University, Van Vleck Observatory, 96 Foss Hill Dr, Middletown, CT 06459}
\author{Kevin Flaherty}\affiliation{Department of Astronomy and Department of Physics\\
Williams College, Williamstown, MA 01267 USA}\affiliation{Wesleyan University, Van Vleck Observatory, 96 Foss Hill Dr, Middletown, CT 06459}
%
%

%

\begin{abstract}


The detection of gas in debris disks raises the question of whether
this gas is a remnant from the primordial protoplanetary phase, or
released by the collision of secondary bodies.  In this paper we
analyze ALMA observations at 1-1.5" resolution of three debris disks
where the \coj \, rotational line was detected: HD~131835, HD~138813,
and HD~156623. We apply the iterative Lucy-Richardson deconvolution
technique to the problem of circumstellar disks to derive disk
geometries and surface brightness distributions of the gas.  The
derived disk parameters are used as input for thermochemical models to
test both primordial and cometary scenarios for the origin of the gas.
We favor a secondary origin for the gas in these disks and find that
the CO gas masses ($\sim 3\times10^{-3}$ \mearth) require production
rates ($\sim 5\times 10^{-7}$ M$_{\oplus}$~yr$^{-1}$) similar to those
estimated for the bona-fide gas rich debris disk $\beta$ Pic.

\end{abstract}

\keywords{open clusters and associations: individual (Scorpius Centaurus) ---
          planetary systems}

\section{Introduction}

Young debris disks trace the final stage of the planet formation
process. Most debris disks have been discovered because of their
excess emission at infrared wavelengths due to orbiting dust
\citep[e.g.][]{aumann1984,Oudmaijer1992,mannings1998}. This dust can
be sustained by collisional cascades involving solids in a wide size
distribution from $\mu$m- to km-sized bodies, triggered by the growth
of Pluto-sized bodies within the disc, by secular or resonant
interactions with planets in the system, and/or by planetesimals born
in high velocity orbits \citep{wyatt2008,wyatt2015}. Imaging of such
systems provides important insights on the architecture of young
planetary systems \citep{hughes2018}.

Observations of gas in debris disks have been more
challenging. Molecular gas was detected in only a few debris disks in
early infrared and millimeter/sub-millimeter observations. This led to
the traditional view that debris disks are gas-poor, and that
therefore the atmospheres of gas- and ice-giant planets must have
formed in the primordial disk phase
\citep{zuckerman95,dent2005,hales2014,moor2015}. With the advent of ALMA,
sensitivity limits improved by several orders of magnitude compared to
earlier surveys and now cold gas in the form of carbon monoxide has been
detected in a growing number of debris disks. Most of them are younger than 50~Myr, with a few exceptions such as Fomalhaut at an age of 440~Myr
\citep{moor2011,Lieman2016,marino2016,kral2017,pericaud2017,moor2017,matra2017b,marino2017}.

The origin of this gas is a matter of ongoing debate. For massive
disks, shielding could prevent the CO from being photo-dissociated and
could explain the persistence of primordial gas even at these advanced
ages \citep[e.g. HD~21997;][]{Kospal13}. These disks have been called
hybrid disks, because their dust content is of secondary origin while
at least a large fraction of the gas could be a remnant from the
protoplanetary disk phase \citep{Kospal13,moor2015,moor2017}.
\citet{kral2018}, however, showed that massive disks can also be
  sustained by cometary collisions if enough atomic carbon (from
  dissociated CO) is accumulated, resulting in a layer that shields CO
  (i.e. shielded secondary disks). In other cases, such as
$\beta$~Pic and HD~181327, low gas densities result in CO lifetimes
shorter than the orbital period and thus the CO must be re-supplied by
collisional activity \citep{Dent2014,marino2016}. In cases where the
observed CO emission is asymmetric, this has been used to argue in
favor of this interpretation \citep{Dent2014,greaves2016,matra2017b}.

The incidence rate of CO gas is significantly higher around
intermediate mass stars than around later spectral
types. \citet{Lieman2016} used ALMA to search for dust continuum and
CO emission toward disks around a sample of 23 B, A, F, and G-type
stars with ages between 11 and 17\,Myr. Three sources were detected in
CO, all corresponding to A type stars, making the detection rate 3/7
for A stars and 0/16 for FGK stars. Newly discovered gas-rich systems
confirm the prevalence of gas around A-type stars.  \cite{moor2017}
estimated an incidence ratio 11/16 around A stars compared to
$~\sim$7\% in FG-type stars. The disks around A stars are also on
average two orders of magnitude brighter than the gas disks around FGK
stars \citep{hughes2018}.

In this work, we focus on the analysis and modelling of the brightest
CO detections in \citet{Lieman2016}: HD~131835, HD~138813, and
HD~156623. Because of their large CO luminosities, they have been
singled out as hybrid disk candidates \citep{moor2017}. A description
of the selection criteria, the full source list, main observational
results and analysis of the continuum data were presented in
\citet{Lieman2016}.  Section~\ref{sample} describes our target
sample. In Section~\ref{modelling} we apply the Lucy-Richardson
deconvolution technique to the case of circumstellar disks to infer
the CO surface brightness distributions.  In Section~\ref{gasmass} we
compare the observations to primordial and cometary origin models for
the gas.  In Section~\ref{discussion} we discuss our results and
Section~\ref{conclusion} presents our conclusions.


\section{CO Rich Debris Disks in Sco-Cen}\label{sample}

\begin{figure}
 \epsscale{0.7}
\plotone{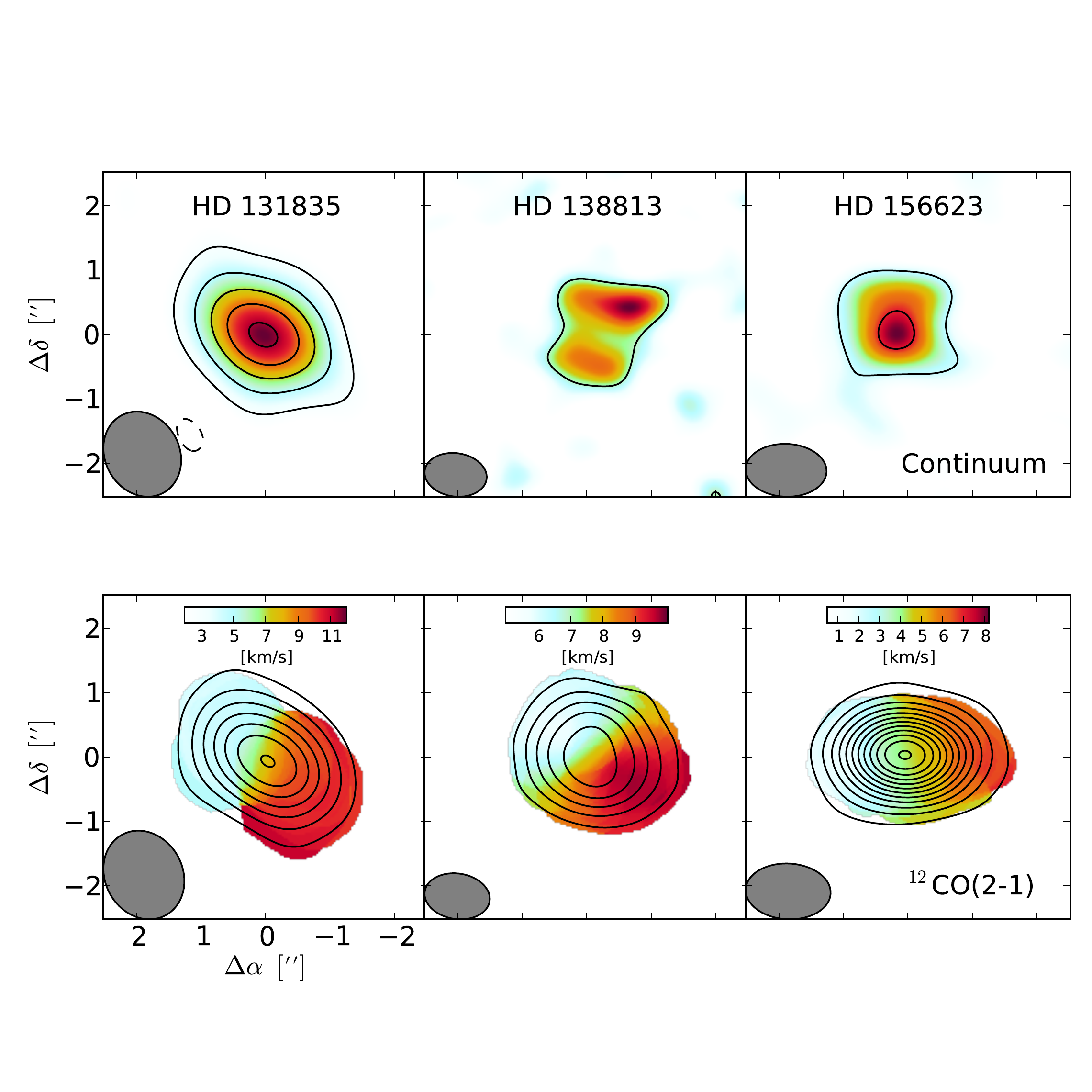}
\caption{ Images of the three debris disks with clear
  \coj\ detections.  Top row: 1.3\,mm continuum images, with contours
  starting at 3$\sigma$ with intervals of 5$\sigma$ \citep[where
    $\sigma$ is 0.079, 0.049 and 0.042~mJy~beam$^{-1}$ (HD~131835,
    HD~138813 and HD~156623 respectively, from ][]{Lieman2016} Bottom
  row: mean \coj\ velocity in \kms\ (color) and \coj\ integrated
  intensity (contours), with contours starting at 5$\sigma$ with
  intervals of 5$\sigma$ The RMS noise in the images and the velocity
  interval to compute the integrated intensity are indicated in
  Table~\ref{tbl:co}.  }
\label{fig:coline2}
\end{figure}

A summary of the ALMA \coj \, observations of 23 luminous debris disks
in the Scorpius-Centaurus Association was presented in
\cite{Lieman2016}.  In this paper we focus on the 3 CO-rich disks that
were detected at sufficient signal-to-noise ratios in order to model
their line intensity profiles (HD~131835, HD~138813, and
HD~156623). Table~\ref{tbl:co} lists the main observational results of
these 3 targets, and Figure~\ref{fig:coline2} shows the resulting
1.3~mm continuum and \coj \, integrated intensity maps after
CLEANing \citep[same as in Figure~1 and Figure~2 from ][]{Lieman2016} . The velocity resolution of the cubes is 0.32\kms, and the
spatial resolution of each image is shown in Table~\ref{tbl:co}.

HD~131835 (HIP~73145) is a well-studied, $\sim$16~My old A4V star
  located in the Upper Centaurus Lupus (UCL) moving group, known to
  harbor a $L_{\rm IR}/L_{*}\sim 2 \times 10^{-3}$ debris disk
  \citep[][]{rizzuto2011,pecaut2012, moor2015}. The dust disk has been
  resolved at near-, mid-infrared, and millimeter wavelengths showing
  the dust extends from $\sim$35~au out to at least 150~au
  \citep{hung2015a,hung2015b,Lieman2016, feldt2017}. The near-IR
  VLT/SPHERE observations resolve at least three sub-structures
  (sub-rings) within the main dust ring, located at 35, 66 and 98~au
  \citep{feldt2017}.  The total dust mass has been derived using
  different methods, providing values ranging from 0.03\mearth to
  7.5\mearth \citep{hung2015b,Lieman2016,feldt2017}, comparable to
  massive debris disks from other surveys
  \citep{roccatagliata2009,thureau2014}.

 



HD~131835 was the only object detected in the $^{12}$CO J=$3-2$ single
dish survey of \cite{moor2015}, out of a sample of 20, 10-40~Myr old,
A- to G-type stars.  \cite{moor2015} showed the $^{12}$CO(3-2) APEX
spectra could be reproduced by a ring-like disk extending from 35~au
to $\sim$120~au radii. The $^{12}$CO(3-2) line is used to derive a
total CO mass of 5.2$\times 10^{-4}$\mearth, assuming the gas is in
local thermodynamic equilibrium (LTE). \citet{moor2017} and \citet{kral2018} derive total CO gas masses of 3.0-6.0$\times 10^{-2}$\mearth\, using observations of the
  optically thin C$^{18}$O(3-2) line. New ALMA observations of
neutral carbon detect 3$\times 10^{-3}$\mearth$\,$ of C$^0$ located
within 40 to 200~au from the star \citep{kral2018}.

%



HD~138813 (HIP~76310) is an A0V star member of the 11~Myr old Upper
Scorpius association \citep{pecaut2012}, host to a $L_{\rm
  IR}/L_{*}\sim 2.1 \times 10^{-4}$ debris disk \citep{dahm2009}. It
was the only A-type star detected in the 1.2~mm single-dish survey of
\cite{mathews2012}, from which they derive a dust mass of
1.1~\mearth. Gas line emission was neither detected with {\it{
    Herschel}} PACS nor JCMT in the survey of \cite{mathews2013}. The
ALMA images are able to resolve a ring-like disk, with inner radius
and width of 70 and 80~AU respectively, and a total dust mass of $\sim
8.3\times 10^{-3}$~\mearth, smaller than the previous estimate from
\cite{mathews2012}. \cite{kral2017} estimate a total CO gas mass of
7.4$\times10^{-4}$~\mearth ~based on the integrated line fluxes from
\cite{Lieman2016} and relaxing the assumption of LTE.

HD~156623 (HIP~84881) is also an A-type star member of the Upper
Scorpius association, with infrared excesses detected in {\it{IRAS}}
and {\it{WISE}} \citep{rizzuto2011}. Prior to the \cite{Lieman2016}
survey, little was known about this star and its circumstellar
environment. The ALMA 1.3~mm images are able to resolve the outer
radius of the disk at $\sim 150$~AU, but do not resolve an inner
edge. \cite{kral2017} estimate a total CO gas mass of
2.0$\times10^{-3}$~\mearth.



\begin{deluxetable}{lcccccc}
\tablecaption{Measured CO J=2-1 Integrated Intensities}
\tablehead{
  \colhead{Source}    &
  \colhead{Beam size} &
  \colhead{P.A.}      &
  \colhead{$\sigma_\mathrm{line}$} &
  \colhead{$\sigma_\mathrm{int}$}  &
  \colhead{$S_\mathrm{CO}$} &
  \colhead{S/N}\\
  \colhead{}                  &
  \colhead{(arcsec)}          &
  \colhead{(deg)}             &
  \colhead{(mJy beam$^{-1}$)} &
  \colhead{(mJy beam$^{-1}$ km s$^{-1}$)} &
  \colhead{(mJy km s$^{-1}$)} &
  \colhead{}
}
\startdata
HD~131835  & $ 1.42\times 1.21$ &  28 &  10.1 &    16 &   798 $\pm$    35 & 22.5\\
HD~138813  & $ 1.02\times 0.71$ &  81 &   7.3 &    14 &  1406 $\pm$    78 & 18.0\\
HD~156623  & $ 1.32\times 0.87$ &  87 &   5.9 &    11 &  1183 $\pm$    37 & 32.3
\enddata
\tablecomments{Summary of ALMA \coj  ~observational results from \cite{Lieman2016}. The columns list
(1) Source name,
(2) full-width-at-half-maximum beam size of the ALMA observations,
(3) position angle of the beam,
(4) rms in the CO J=2-1 spectral images per 0.32\kms\ channel,
(5) rms in the CO J=2-1 integrated intensity images measured in an annulus between
4 and 8\arcsec\ centered on the stellar position. The CO was integrated between velocities
of 3 and 12\kms\ (HD~131835), 
5 and 10.6\kms\ (HD~138813), and 0.5 and 8.2\kms\ (HD~156623).
(6) Integrated CO J=2-1 intensity measured in the ALMA images. An aperture radius of 
2\arcsec\ was used for HD~131835, HD~138813, and HD~156623.
(7) Signal to noise ratio of the measured CO integrated intensity.
}
\label{tbl:co}
\end{deluxetable}

\section{Modeling}\label{modelling}

In order to investigate the possible origins of the CO gas, we must
first characterize its spatial location within the disk. In this
section we implement a method that allows to derive the surface
brightness distribution of the gas in spatial scales smaller than the
angular resolution, by taking advantage of the {\textit{a priori}}
knowledge of the gas kinematics. This information is generally
available in many astrophysical problems (e.g. galactic rotation
curves), and in the particular case of circumstellar disks the gas can
be well described by a keplerian velocity field
\citep{hughes2008,Kospal13,Dent2014}.

\subsection{Lucy-Richardson Deconvolution}\label{lucy}

The `Lucy-Richardson Deconvolution' is an iterative rectification
method for observed distributions, presented independently by
\citet{richardson1972} and \citet{lucy1974}. This method attempts to
restore the {\textit{original}} distribution from an
{\textit{observed}} distribution, in which the observed distribution
corresponds to a {\textit{degraded}} version of the original
distribution. The key to this method is that the original distribution
can be recovered if sufficient {\textit{a priori}} information about
the degrading process is available. Examples of degrading processes
are the spatial distortion by an instrumental point spread function
(PSF), instrumental broadening of spectral lines (equivalent to a
velocity PSF), additive noise, or other more complex processes
\citep[e.g.,][]{zech2013,stock2015,zorec2016}.

The use of {\textit{ a priori }} kinematical information for deriving
CO emissivity distribution in galaxies was demonstrated over twenty
years ago by \citet{scoville1983}. By applying the \citet{lucy1974}
iterative rectification scheme, \citet{scoville1983} derived a
de-projected emissivity distribution that  successfully reproduced
the observed line intensity profiles (in their case, the velocity
field of the galaxy was known {\textit{a priori}} via optical line
studies). The advantage of this method is that it does not need any
previous assumption on the surface brightness distribution
(e.g. whether it is a power-law or other functional form), and that it
converges relatively fast (typically in less than ten
iterations).

In the case of astrophysical disks (circumstellar or galactic), the
observed line intensity profiles result from the (double) convolution
between the spatial and velocity PSFs, with the intrinsic emissivity
distribution. If $x$ and $y$ are the linear displacement coordinates
parallel and perpendicular to the disk's major axis, we consider a
point {\it{(x,y)}} located at distance R from the center of a circumstellar
disk. If the velocity field is known at all points (e.g. Keplerian),
$\rho(R)$ is the axisymmetric emissivity distribution and $i$ the disk
inclination angle, then in the case where there is no instrumental
broadening the measured intensity in that pencil beam point is:

\begin{equation}\label{eqn1}
  J(x,y;v)=\frac{\rho(R)}{\cos i}\delta(v-v_{x,y})
\end{equation}

\noindent where $v_{x,y}$ is the Keplerian velocity at position
{\it{(x,y)}}. In reality, the observed line intensity profiles correspond to
convolution of the emissivity distribution $\rho(R)$ with the
instrumental PSF:

\begin{equation}\label{eqn2}
  I(x,y;v) = \int \int \int J(\epsilon,\eta,w)\,P(x,y,v|\epsilon,\eta,w)\,d\epsilon\,d\eta\,dw.
\end{equation}

\noindent Converting integration variables to polar coordinates
($\epsilon=R\cos\theta $ and $\eta~=~R\sin\theta\cos~i$),
Equation~\ref{eqn2} can be written as

\begin{equation}\label{eqn3}
\resizebox{0.8\textwidth}{!}{$  I(x,y;v) = \int \int \int J(\epsilon,\eta,w)\,P(x,y,v|R\cos \theta,R\sin \theta \cos i,v_{R,\theta})\,R\,dR\,d\theta $}
\end{equation}

\noindent The instrumental PSF consists of two terms, the spatial PSF $P_s$ and
the velocity spread function $P_v$. The spatial component of the
instrumental PSF is written as

\begin{equation}\label{eqn4}
P_s(x,y|\epsilon,\eta)=\frac{1}{2\pi\sigma_s^2}\exp\left[\frac{(x-\epsilon)^2+(y-\eta)^2}{2\sigma_s^2}\right],
\end{equation}

\noindent where $\sigma_s^2$ is computed taking into account the FWHM of the
minor and major axis of the observed PSF, as well as its positions
angle in the sky. The velocity-broadening PSF takes the form:

\begin{equation}\label{eqn5}
P_v(v|w)=\frac{1}{\sqrt{2\pi\sigma_v}}\exp\left[\frac{(v-w)^2}{2\sigma_v}\right]
\end{equation}

\noindent The dispersion of the velocity PSF, $\sigma_v$,  is given by the thermal and
non-thermal line widths added in quadrature
\citep[e.g.][]{Hughes2010},\\

\begin{equation}\label{eqn6}
\sigma_v(r)=\sqrt{\frac{2k_BT(r)}{m} + \xi^2},
\end{equation}

\noindent where $k_B$ is Boltzmann's constant, $T(r)$ is the local disk
temperature, $m$ is the average mass per particle, and $\xi$ is a
velocity broadening term adjusted to match the instrumental spectral
resolution. No turbulence has been assumed.  The local disk
temperature $T(r)$ is computed assuming that $T_{gas}(r)=T_{dust}(r)$
and that the dust grains are in thermal equilibrium with the stellar
radiation, i.e:

\begin{equation}\label{eqn7}
T(r)=\Big(\frac{L_*}{16\pi \sigma R^2}\Big)^{1/4},
\end{equation}

\noindent where $L_*$ is the stellar luminosity and $\sigma$ is the
Stefan-–Boltzmann constant.  \noindent The velocity broadening due to Keplerian rotation is estimated by
assuming a geometrically flat, azimuthally symmetric circumstellar
disk viewed at inclination angle $i$, in polar coordinates

\begin{equation}\label{eqn8}
v_{r,\theta}=\sqrt{\frac{GM_*}{R}}\cos \theta \sin i . 
\end{equation}

\noindent The double convolution Kernel $\Pi(x_k,y_k;v_l|R_j)$ is defined as the
convolution between the spatial PSF $P_s$ and the velocity spread
function $P_v$ \citep{scoville1983}, \\

\begin{equation}\label{eqn9}
\resizebox{0.7\textwidth}{!}{$ \Pi(x,y;v|R)=\frac{1}{2\pi}\int_0^{2\pi}P_s(x,y|R\cos \theta,R\sin \theta \cos i)P_v(v|v_{R,\theta})d\theta.$}
\end{equation}

\noindent By combining (\ref{eqn1}), (\ref{eqn3}) and (\ref{eqn9}), the
predicted intensity profile at any given position can be calculated
from

\begin{equation}\label{eqn10}
I^n(x_k,y_k;v_l)=2\pi\sum_{J=1}^{n_R} R_J \rho^n(R_J)\Pi(x_k,y_k;v_l|R_j)\Delta R.
\end{equation}

\noindent Note that the $I^n(x_k,y_k;v_l)$ is integrated over an area
corresponding to the instrumental beam.  \citet{scoville1983} showed
that for an axisymmetric disk, the de-projected emissivity distribution
$\rho(r)$, can be derived on scales much finer than the instrumental
spatial resolution by iterating:

\begin{equation}\label{eqn11}
\rho^{n+1}(R)=\rho^n(R)\frac{\sum_{k=1}^{n_B} \sum_{l=1}^{n_v} \left [ \frac{\widetilde{I}(x_k,y_k;v_l)}{I^n(x_k,y_k;v_l)}\right ] \Pi(x_k,y_k;v_l|R_j)}{\sum_{k=1}^{n_B}\sum_{l=1}^{n_v} \Pi(x_k,y_k;v_l|R_j)},
\end{equation}

\noindent where $\widetilde{I}(x_k,y_k;v_l)$ is the observed intensity at
position $(x_k,y_k)$ and velocity v$_l$, and $I^n(x_k,y_k;v_l)$ is the
theoretical line intensity profile that would be observed given the
surface brightness distribution $\rho^n(r)$. The positions and
velocities $x_k$, $y_k$ and $v_l$ sampled at a total of $n_B$
positions and $n_v$ channels, respectively. The radial emissivity
distribution $\rho(r)$ is computed at $n_R$ discrete points spaced by
$\Delta R$. The separation $\Delta R$ can be much finer than the
nominal spatial resolution.


Starting with a constant surface brightness distribution
$\rho^0(r)=1$, the theoretical line intensity profile is calculated
using Equation~\ref{eqn10} and $\rho^0$ at $n_B$ positions, and the
reduced $\chi^2$ between the observed and theoretical line intensity
profiles is computed as

\begin{equation}\label{eqn12}
\chi^2=\frac{\sum_{j=1}^{n_B}\sum_{v=1}^{n_v}(\widetilde{I}_{j,v} - I^n_{j,v})^2}{\sigma^2},
\end{equation}




\noindent where $\sigma$ corresponds to the RMS noise per channel in the
observations. We adopt a grid that samples every $1/3$ of the beam to
prevent over-sampling and avoid excessive computational time. For all
sources the sampled area is limited to a circular region centered at
the stellar position of 2 arcseconds in radius. \\

\begin{figure}
\begin{center}
\epsscale{0.5}
\includegraphics[angle=0,scale=0.4]{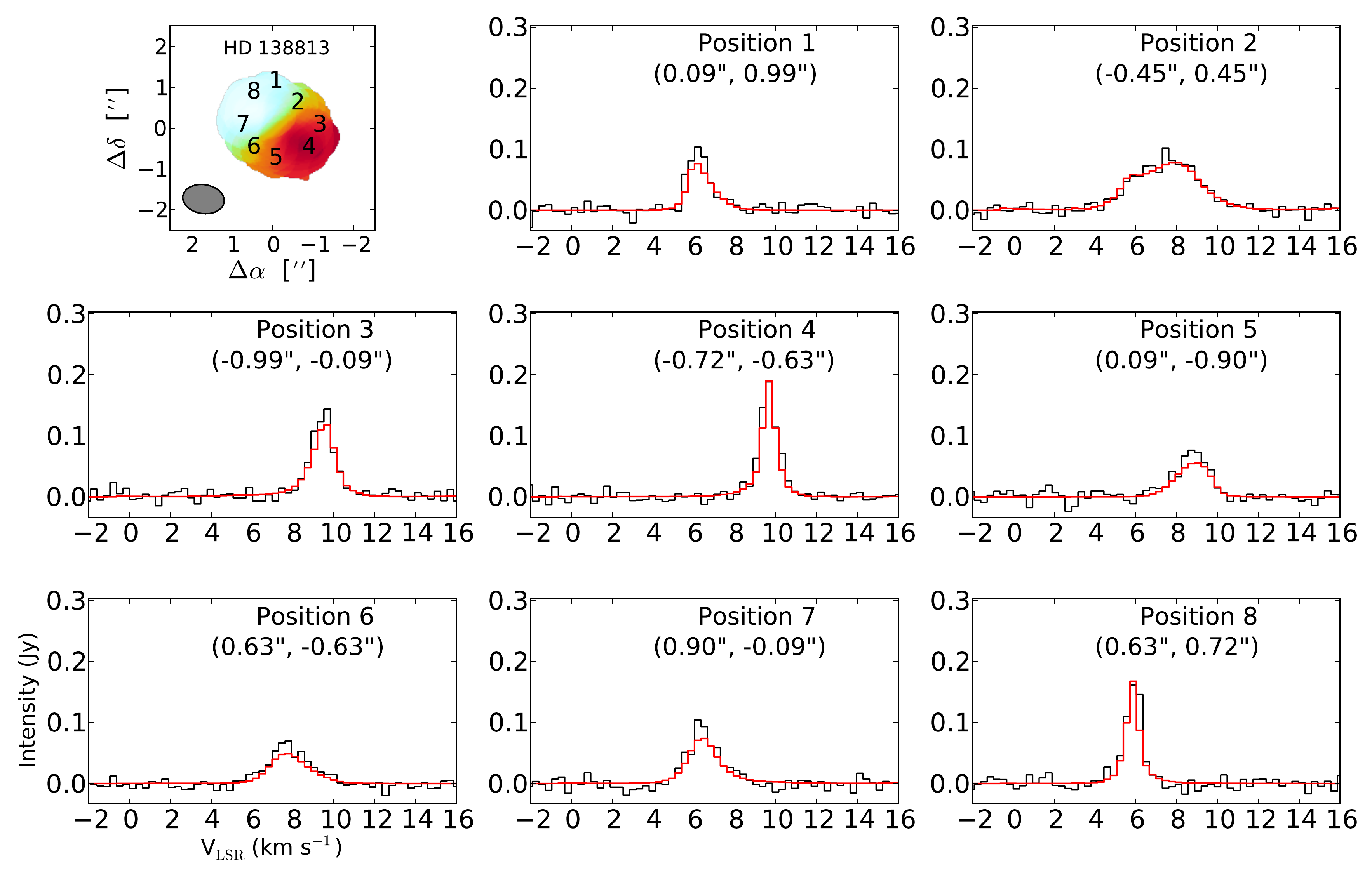}
\caption{Observed $I(x,y;v)$ versus model $I^n(x,y;v)$ line intensity
  velocity profiles for HD~138813 (black and red histograms
  respectively). The different panels show the \coj\ velocity
    profiles at different $(x,y)$ coordinates with respect to the
    disk's center (the corresponding $(x,y)$ position offsets are
    listed in each panel). The top-left panel shows the moment 1 map
    with the position of the different spectra. Only 8 out of the
    total 172 spectra which were extracted from $n_B$ different
    positions are shown.  The model shown corresponds to the best fit
    model.}  
\label{fig:lineprofiles}
\end{center}
\end{figure}

We iterate Equation~\ref{eqn11} until the reduced $\chi^2$ reaches
  a value less than unity or when the fractional change in the reduced
  $\chi^2$ is less than 1$\%$ (which was found to be a good criteria for convergence of the parameters).
This approach allows to derive quickly
(generally less than 10 iterations) the surface brightness
distribution of the disk that provides the best fit to the models in
comparison to other more time-consuming methods \citep[e.g. radiative
  transfer modeling and/or visibility
  computation;][]{isella2009,tazzari2016}. In
Figure~\ref{fig:lineprofiles} we show an example of the observed
$\widetilde{I}(x,y;v)$ and modeled $I^n(x,y;v)$ line profiles for the
best-fit model for HD~138813 (see Section~\ref{modelling}).

\subsection{Disk modeling and parameter-space search}

In our modeling approach we adopt a Bayesian method to obtain
probability distribution functions of the model free parameters.  The
parameter search is performed using the Python package
{\textit{emcee}} \citep{foreman2013}, which uses the affine-invariant
implementation of MCMC to run simultaneously several Markov chains to
map the posterior probability distribution. By using many walkers, and
by proposing new models based on the relative positioning of the
{\textit{walkers}}, {\textit{emcee}} is effective at handling posterior
  probability density functions (PDF) with strong degeneracies.


\begin{deluxetable}{lccc}
\tablecaption{Fixed Model Parameters}
\tablehead{
  \colhead{} &
  \colhead{HD~131835} &
  \colhead{HD~138813} &
  \colhead{HD~156623} 
}
\startdata
Stellar parameters &          &     &    \\ 
 \hline    
L$_{\star}$    (L$_{\odot}$) $^{a}$ & 11.0     &   20.4    &  13.3  \\
M$_{\star}$    (M$_{\odot}$) $^{b}$ & 1.77     &   2.2     &  2.2   \\
Distance (pc)$^{c}$      & 133.7    &  137.5    & 111.8  \\
                           &          &           &    \\ 
\hline 
Spectra                    &          &            &    \\ 
\hline 
$n_B$      &             71  &          172   &     113  \\ 			
$n_v$      &             33  &           21   &      35  \\ 		
v$_1$   (\kms)  $^{d}$       &   2.0  &         4.6   &    -1.0  \\ 			
v$_2$   (\kms)  $^{d}$       &  12.0  &        11.0   &    10.0  \\ 				
dv   (\kms)          &   0.32 &         0.32  &     0.32  \\
$\Delta R$   (arcsec) $^{e}$ &   0.1  &         0.1   &     0.1  \\ 		
\enddata

\tablecomments {$^{(a)}$ Luminosities are derived using the values in
  \citet{kral2017} scaled by the difference in distance between
  Hipparcos and Gaia DR2.  $^{(b)}$ Stellar mass for HD~131835 is taken
  from \citet{moor2015}, and from \citet{hernandez2005} for
  HD~138813. For HD~156623 the same mass as HD~138813 is
  assumed. $^{(c)}$ Distances are obtained from the second data release (DR2) of Gaia \cite[]{gaia2018}.   $^{(d)}$ v$_1$ and v$_2$ denote the velocity range used for   the fitting. $^{(e)}$  $\Delta R$ corresponds to the radial sampling of the
  surface brightness distribution.}
\label{tbl:fix}
\end{deluxetable}

Our disk model is defined by the disk's position angle
{\textit{$\theta$}}, inclination {\textit{i}}, central velocity
{\textit{v$_r$}}, as well as the gas surface brightness distribution
$\rho(r)$. In addition two extra parameters, {\textit{$\Delta\alpha$}}
  and {\textit{$\Delta\delta$}}, are added to find the exact centroid
    position of the disk \citep[e.g.][]{tazzari2016}.  Table~\ref{tbl:fix} shows the fixed model parameters for each system.

\begin{figure}
\epsscale{0.75}
\plotone{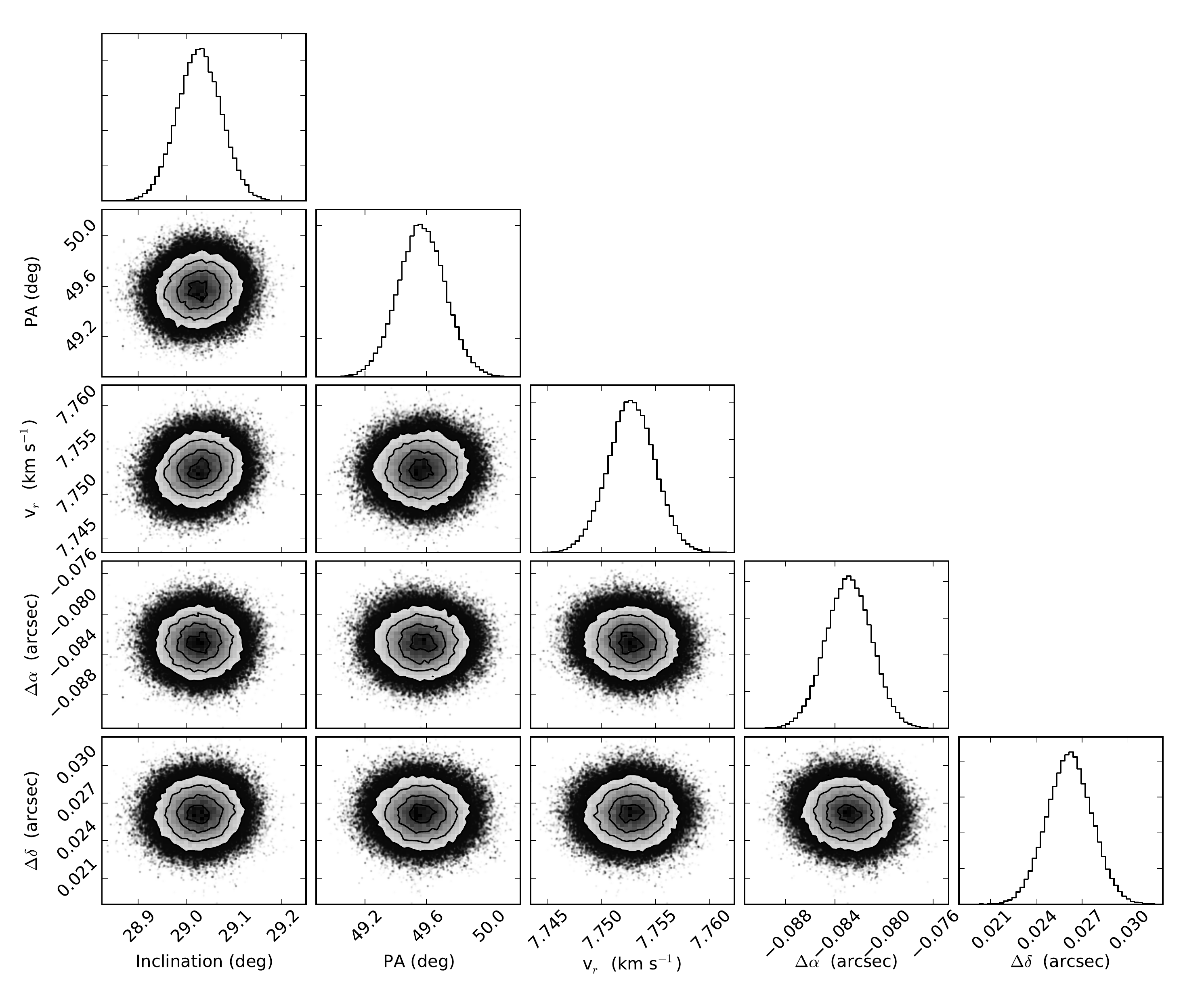}
\caption{Results of the MCMC for HD~138813, showing the one and two
  dimensional projections of the posterior probability, after skipping
   the first 200 steps.  }
\label{fig_corner_76310}
\end{figure}

At each iteration, the position of a given walker in the parameter
space (i.e. one set of model parameters) defines a disk model, for
which the best-fit surface brightness distribution is found by
Lucy-Richardson deconvolution as described in \ref{lucy}. The $\chi^2$
of this model is computed using its surface brightness distribution,
and the resultant $\chi^2$ value is used to compute the PDF. The next
iteration will consider the PDF computed by all walkers in order to
define the move to the next point in parameter space. In this manner,
the walkers interact with each other and the PDF can be sampled fast
and efficiently.

Typically the MCMC chains are let to evolve during the
{\textit{burn-in} phase, during which 1000 walkers sample a broad
  range of the parameter space in order to locate the maximum of the
  posterior probability. This is achieved in between 100 to 400
  steps. After the burn-in phase, 800 to 1000 more iterations are
  allowed in order to further sample the posterior probability around
  the global maximum. After the removal of the burn-in steps, the
  posterior probability provides information on the marginal
  distributions of the free parameters (i.e. the uncertainties on the
  derived free parameters). The advantage of this method is that it
  can be fully parallelized. Using 24 processors in parallel we were
  able to sample 1200 iterations of the 1000 individual chains in a
  week of processing time.

\begin{figure}
\epsscale{0.41}
\plotone{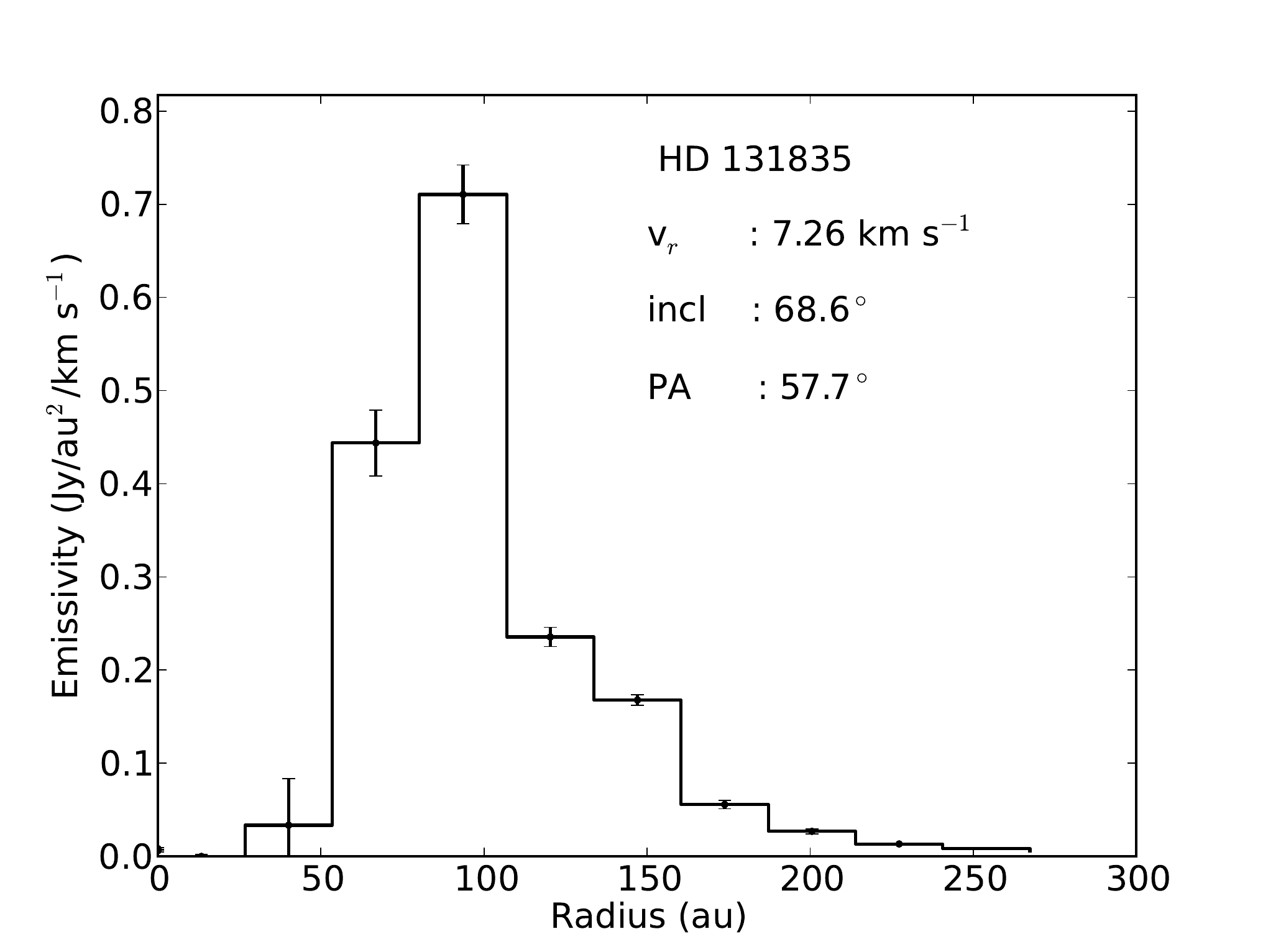}\hspace{-0.8cm}
\plotone{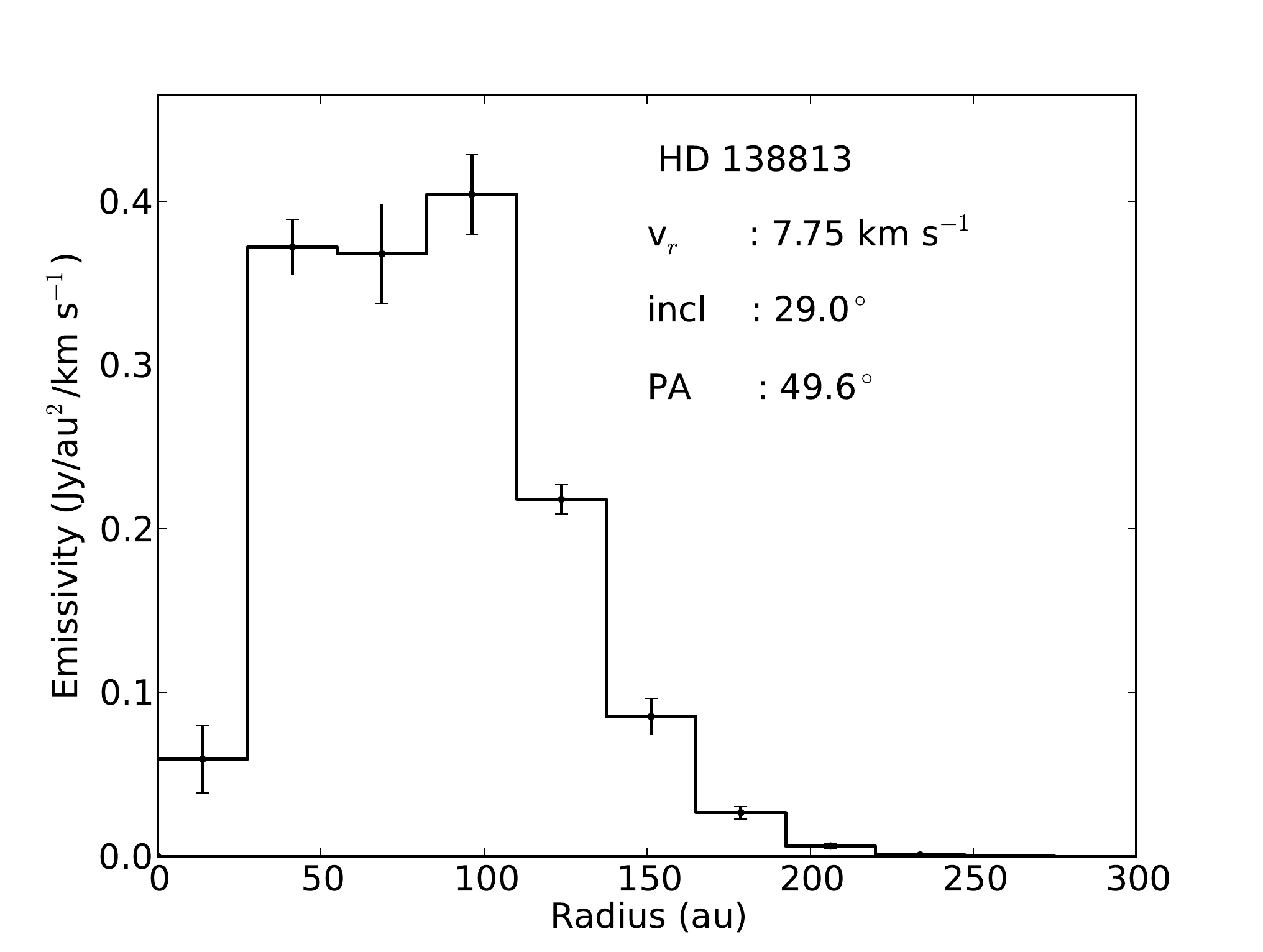}\hspace{-0.8cm}
\plotone{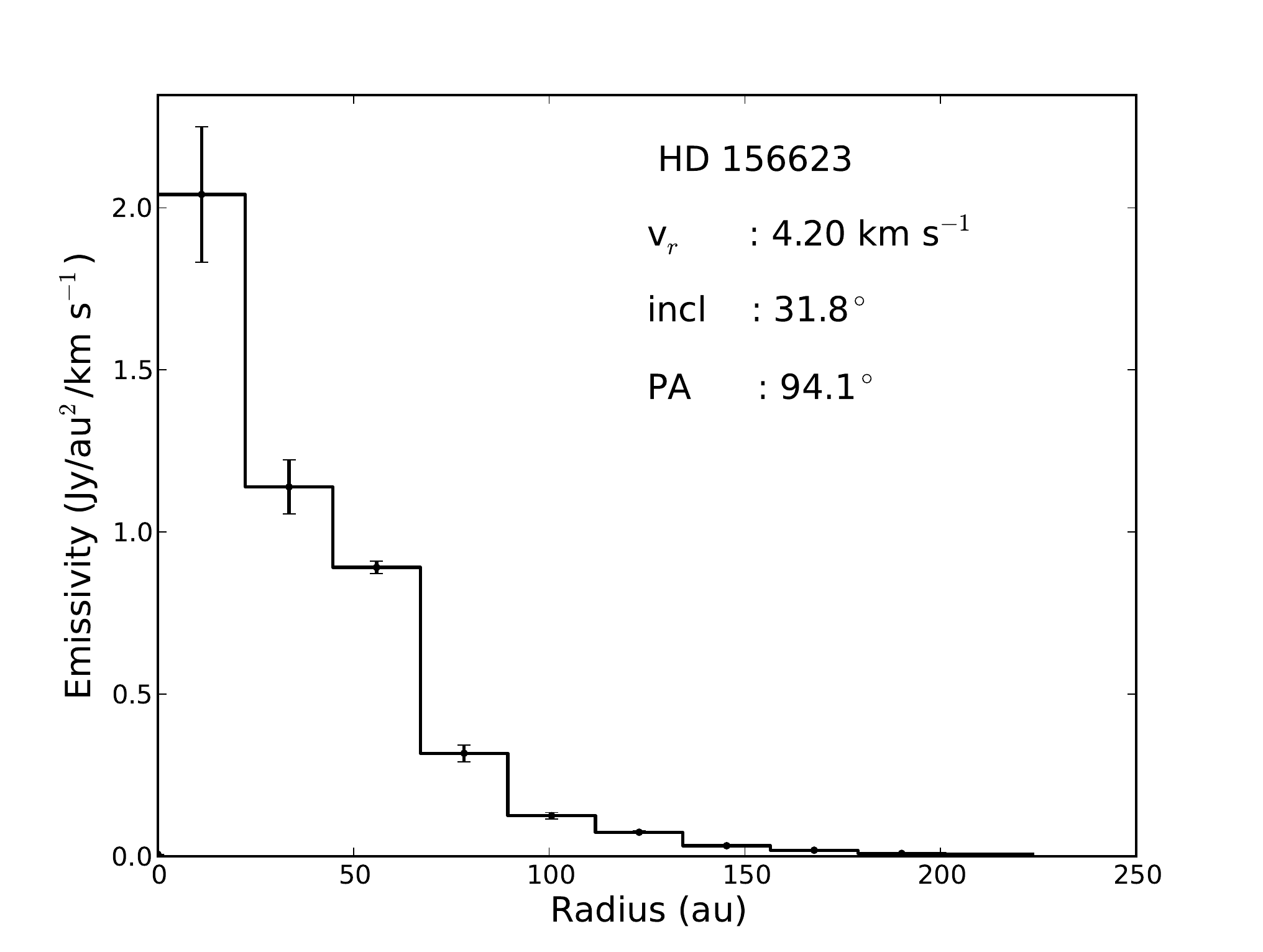}
\caption{Derived CO emissivity distribution for HD~131835 (Left),
  HD~138813 (Center) and HD~156623 (Right). The uncertainties in
    each emissivity bin were estimated by taking the standard
    deviation between 100 different random samples of the PDF.} 
\label{fig:emissivity}
\end{figure}

\subsection{Modeling results}\label{modelling}

The MCMC search through Lucy-Richardson deconvolution model fitting
method was run for each of the three gas
disks. Figure~\ref{fig_corner_76310} shows the staircase plots of the
chains obtained for HD~138813 after the MCMC fitting process. The MCMC
staircase plots for HD~131835 and HD~156623 are presented in
Section~\ref{lucyappendix} of the Appendix. In Table~\ref{tbl:mod} we
show the best-fit parameters for each disk. The uncertainties on the
individual parameters are estimated by fitting a Gaussian to the
marginalized distributions (corresponding to the diagonal panels of
the staircase plots). Figure~\ref{fig:emissivity} shows the CO surface
brightness distribution computed from the best-fit model of each disk.

Figure~\ref{fig:hd76310-residual} shows the comparison between the
data and model for the HD~138813 disk, together with the resulting
residuals in both integrated intensity (moment 0) and velocity
dispersion (moment 1) maps.  It can be seen from the residual maps
that the models provide reasonable fit to the observations, with no
residuals above 5-$\sigma$ level (15\% of the peak flux). Data,
model and residual plots for HD~131835 and HD~156623 are presented in
Section~\ref{lucyappendix} of the Appendix.


\begin{figure}
\epsscale{0.9}
\plotone{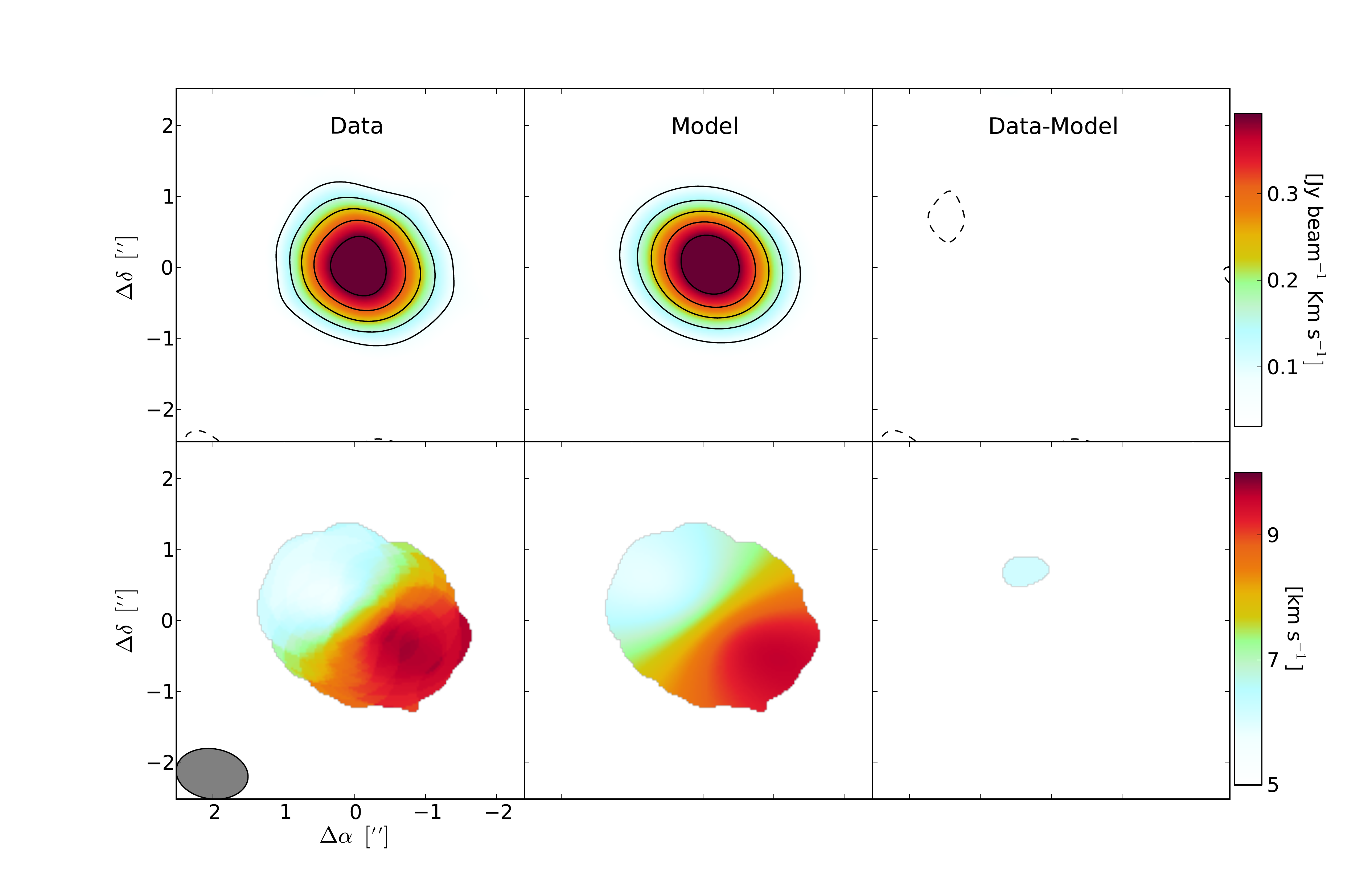}

\caption{
    {\it{Top}}: Integrated intensity map for the data, model and residuals of the  CO emission of the HD~138813 disk. Contours start at 5$\sigma$ with intervals of 5$\sigma$. Negative contours  start at -2$\sigma$ with intervals of -5$\sigma$ (dashed lines). 
 {\it{Bottom }}: Intensity-weighted mean velocity (moment 1) for the data, model and residuals of the  CO emission of the HD~138813 disk. 
}
\label{fig:hd76310-residual}
\end{figure}

\begin{deluxetable}{lccccc}
\tablecaption{Best-fit Model Parameters}
\tablehead{
  \colhead{Source}    &
  \colhead{Inclination}  &
  \colhead{PA ($^{\circ}$) } &
  \colhead{v$_r$} &
  \colhead{$\Delta\alpha$}&
  \colhead{$\Delta\delta$}\\
  \colhead{} &
  \colhead{(deg)} &
  \colhead{(deg)} &
  \colhead{(km s$^{-1}$)} &
  \colhead{(arcsec)} &
  \colhead{(arcsec)} 
}
\startdata
HD~131835    & 68.6 $\pm$1.6&  57.7 $\pm$6.2 & 7.26  $\pm$0.06 & 0.03$\pm$0.06 &-0.02$\pm$0.06 \\
HD~138813    & 29.0 $\pm$0.3&  49.6 $\pm$0.8 & 7.75  $\pm$0.01 & -0.08$\pm$0.01 & 0.03$\pm$0.01 \\
HD~156623    & 31.8 $\pm$0.5&  94.1 $\pm$2.8 & 4.20  $\pm$0.03 &  0.05$\pm$0.02 & 0.01$\pm$0.02 
\enddata

\tablecomments{The uncertainties in the derived parameters correspond to 1$\sigma$ and are obtained by a fitting a Gaussian to the marginalized distributions, after correcting for the number of correlated  pixels within one beam. 
}
\label{tbl:mod}
\end{deluxetable}

\begin{figure}
\centering
\includegraphics[scale=0.4]{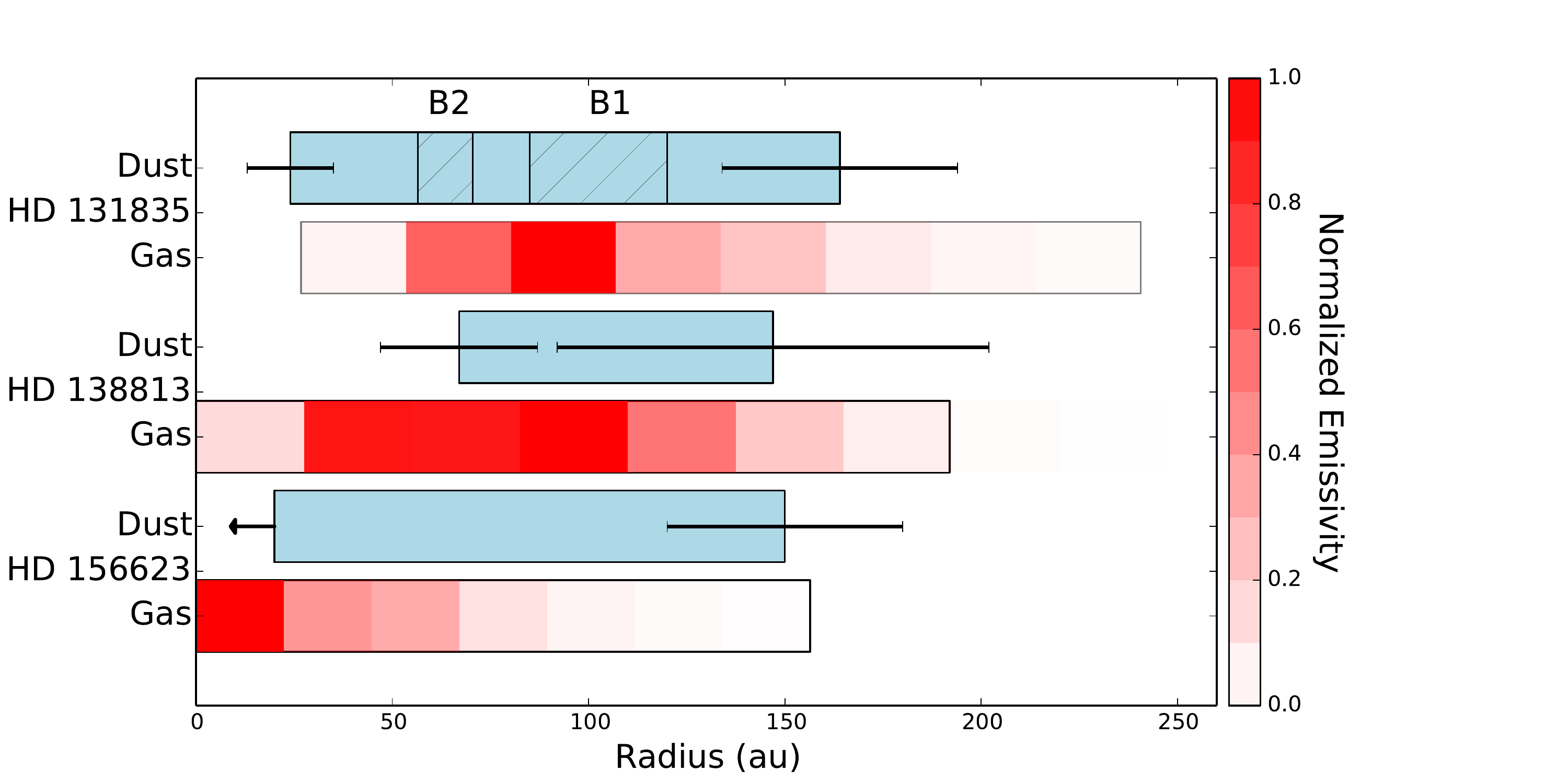}
\caption{ Graphical representation dust disk radii \citep[from
  ][]{Lieman2016} versus the gas disk radii derived in this work. The
  colorscale shows the CO emissivities from
  Figure~\ref{fig:emissivity}, which have been normalized to unity.
  The location of the two bright near-IR dust rings detected with
  SPHERE around HD~131835 are also shown (B1 and B2 respectively).
  }
\label{diskradii}
\end{figure}

 HD~131835 is the most studied of the targets within our sample.
 \cite{moor2015} was able to model the $^{12}$CO(3-2) APEX spectra
 assuming the gas extends from 35~au \citep[obtained from the
   continuous disk model in ][and consistent with the location of the
   first inner ring resolved with SPHERE]{hung2015b}, and varying the
 outer radius to fit the single dish data.  For their best fit model,
 \cite{moor2015} measure a systemic velocity of 7.2~km~s$^{-1}$
 in the Local Standard of Rest (LSR) frame. The systemic
 velocity we derive is identical to the one measured by
 \cite{moor2015}. This value of the system's velocity is also
 consistent with the stellar radial velocity measured in the optical
 by \cite{Rebollido}. They measure a heliocentric radial velocity of
 v$_{\rm Helio}=$2.6$\pm$1.4~km~s$^{-1}$, which corresponds
 6.4$\pm$1.4~km~s$^{-1}$ after converting to v$_{\rm LSR}$. This is
 also consistent with the estimation of the systemic velocity from the
 [CI] line data by \citet{kral2018}, who derive a heliocentric
 velocity of 3.57$\pm$0.1 ~km~s$^{-1}$, which corresponds to a v$_{\rm
   LSR}$ of 7.37 $\pm$0.1~km~s$^{-1}$. The radial velocities derived
 for HD~138813 and HD~156623 are also consistent with radial
 velocities measured in the optical. \cite{Rebollido} find radial
 velocities of 7.7$\pm$2.0~km~s$^{-1}$ and 4.6 $\pm$1.5~km~s$^{-1}$
 for HD~138813 and HD~156623 respectively (v$_{\rm LSR}$).

 The inclination and position angles adopted in the models from
 \cite{moor2015} were fixed to the values derived from Gemini/GPI
 observations \citep[74$^{\circ}$ and 50$^{\circ}$
   respectively;][]{hung2015b}. More recently \cite{feldt2017} refined
 the inclination and PA of the near-IR dust disk to
 72$^{\circ}$.6$\pm_{-0.6}^{0.5}$ and
 60$^{\circ}$.3$\pm_{-0.2}^{0.2}$, respectively. \citet{kral2018} fit
 an inclination of 76.95$^{3.1}_{2.4}$ to the [CI] line data.  The
 inclination angle of 68.6 $\pm$1.6 we derive is consistent within
 1.7$\sigma$ with the values from the infrared images and also from
 the ALMA CI data \citep{kral2018}. For comparison, we compute the
 inclination of the 1.3mm continuum dust disk by performing a gaussian
 fitting of the 1.3mm data. Using {\sc CASA} task {\tt imfit} to fit
 in the image plane we derive an inclination angle of 69.8$\pm$0.7,
 whereas fitting in the visibility domain we obtain an inclination
 angle of 66.4$\pm$0.2 (Using {\sc CASA} task {\tt uvmodelfit}).  The
 scatter seen in the continuum inclination could be indicative of the
 limitations of the data or that there are systematics affecting the
 derivation on the inclination angle and thus the formal errors quoted
 for the inclination of the gas disk could be
 underestimated. Differences between the inclinations derived by
 millimeter and scattered light images are not
 uncommon. \cite{loomis2017} reported differences of
 $\sim$15$^{\circ}$ in the disk around AA~Tau, which could be
 attributed to a warp in the inner regions of the disk. Discrepancies
 in the inclinations of the gas and dust disks measured in the
 millimeter have also been reported, as is the case of $\beta$
 Pictoris in which the CO is more clumpy than the dust and is located
 5~au above the midplane more closely aligned with an inner disk and
 to the orbit of the planet $\beta$ Pic $b$ \citep{matra2017a}.  There
 are no previous measurements of the disk inclinations for the
 HD~138813 and HD~156623 disks. Higher resolution images of the HD~131835 system are necessary to
investigate the presence of clumps, warps or other asymmetries in the
gas disk. 

It is interesting to note that similar to the inner and
outer radii used by \cite{moor2015} to fit the $^{12}$CO(3-2) APEX
data, our modelling of the $^{12}$CO(2-1) line derives a surface
brightness distribution which is also confined to a region between
50-150~au. This is also similar to the extension of the CI disk
\citep[40 to 200~au][]{kral2018}.  To quantify how much the assumption
of T$_{\tt gas}$=T$_{\tt dust}$ could affect the determination of the
emissivity distribution we experimented using a temperature profile
(T(r) $\propto$ r$^{-p}$ with p=0.4) that yields temperatures
of $\sim$100~K in the inner 20~au, similar to the ones used in the
models of \cite{kral2017}. We find that this does not affect the
determination of a cavity in the CO, while the derived disk parameters
remain consistent within 2-3 sigma.

Figure~\ref{diskradii} shows the comparison between the location of
the dust and gas disks for all three targets. The inner and outer
radii of the dust emission are taken from \citet{Lieman2016}, while
the location of the gas correspond to the emissivity profiles from
Figure~\ref{fig:emissivity}.  The dust and gas ring around HD~131835
are both confined to a $\sim$100~au ring. In Figure~\ref{diskradii} we
have also marked the location of the two brightest dust rings seen in
the near-IR with SPHERE \citep[named B1 and B2 by][]{feldt2017}. The
two near-IR dust rings roughly coincide with the peak of the CO
emission, which could suggest a common origin. Similarly to
  HD~131835, HD~138813 shows a ring-like structure in dust
  \citep{Lieman2016}, and also in $^{12}$CO according to our modelling
  of the ALMA data. The CO surface brightness from HD~156623 is found
to be centrally peaked. Since the inner radius of the dust disk was
unresolved at 1.3~mm, thus it is not possible to conclude whether the
peak gas emission resides inside a dust cavity or if it is mixed with
the dust. Most of the gas (20 to 60$\%$ of the peak surface
brightness), however, appears to be co-located with the dust.

\section{Gas Mass and Origin}\label{gasmass}

We next attempt to determine the mass of CO implied by the observed
emission. We first present simple estimates and then discuss results
obtained using detailed thermochemical models where the temperature,
density and chemical structure are calculated in a self-consistent
manner by solving for hydrostatic pressure equilibrium coupled with
thermal balance.  We consider both proposed origins of the gas, one in
which the gas is a primordial relic from the protoplanetary disk
stage, and another where the gas is produced by secondary collisions
in a debris disk
\citep[e.g.][]{moor2011,moor2017,zuckerman2012,kral2017}.

\paragraph{Estimate of optically thin LTE mass limit} A simple CO mass estimate can be made by assuming that the emission is
in LTE and optically thin. The fractional population of the J=2 level
is $N_{2}/N_{CO} = (2J+1) e^{-16.6/T}/{Z(T)}$ where $Z(T)~\sim
T/2.762$ is the rotational partition function for CO
\citep{Hollenbach1979}. For a mean line luminosity of $\sim
  10^{-9}$ L$_{\odot}$ as observed (a flux of 1~Jy~km~s$^{-1}$ from a
  disk at 100~pc gives a line luminosity
  $\sim 2\times\,10^{-9}$~L$_{\odot}$), and with a transition probability
$A_{21}=6.91\times10^{-7}$ s$^{-1}$, we can estimate the total number
of CO molecules from the total line luminosity $L_{21} = N_{2} A_{21}
\Delta E_{21}$ giving a CO mass

\begin{equation}
{\rm M}_{CO}^{LTE, \tau<1} = 1.25 \times
10^{22}\ (\frac{L_{21}}{10^{-9}{\rm L}_{\odot}}) \ T \ e^{16.6/T}
\quad g
\label{mcothin}
\end{equation}
and a minimum mass M$_{CO}^{min}$ of $5.6\times10^{23}$ g or $10^{-4}$
M$_{\oplus}$ if all the gas is at $\sim 16.6$K (corresponding to the
upper energy level of the CO(2-1) line).

Deviating from any of the assumptions made above on the optical depth,
LTE conditions or temperature would result in a mass higher than
M$_{CO}^{min}$ to reproduce the observed emission.  From
Eq.~\ref{mcothin}, the mass derived is seen to increase for gas
temperatures both higher and lower than 16.6K. ($\sim 3$M$_{CO}^{min}$
for 5~K and $\sim 5$M$_{CO}^{min}$ for 200K).

\paragraph{Estimate of mass required for LTE}

The fact that we have spatially resolved emission maps allows us to
estimate a density from the mass and place constraints on the validity
of the LTE assumption. For the two scenarios considered, we assume
that H$_2$ is the main collision partner for primordial gas and that
electrons are the main colliders for secondary origin gas (H atoms
from photodissociation of water could be abundant, but these collision
rates are lower and electrons dominate, e.g., see
\citet{matra2017a}. For thermally populated levels, the collider gas
density needs to be higher than the critical density
($n_{crit}=A/\gamma \sim 10^4$ cm$^{-3}$ for H$_2$ \citep{yang2010},
and $\sim$100 cm$^{-3}$ for e$^-$ \citep{Dickinson1977}) everywhere in
the disk.

From the radial intensity distribution fits (Section~\ref{modelling},
Figure~\ref{fig:emissivity}) and from the emission maps, the CO disk
radius is $\sim$ 200~au. Assuming a constant disk thickness of
10~au \citep[similar to the Kuiper belt;][]{jewitt1996,trujillo2001},
the disk volume is $\sim 10^6$ au$^3$.

The mass needed for LTE can be estimated with a simple constant
density assumption as $\sim n_{crit} \times $volume; the primordial
origin scenario therefore requires a CO disk mass of $\sim 2\times
10^{23}$g (for $n(CO)\sim 1.4\times 10^{-4} n(H_2)$). This limit is
similar to that made with the optically thin assumption above, and LTE
is likely a valid assumption for the primordial gas disk.  The total
disk mass in this case (H$_2$+CO) is expected to be at least $\sim
0.01$ M$_{\oplus}$. In the case of secondary origin gas,
  densities are low enough that LTE may be difficult to
  attain. Emission is therefore likely to be sub-thermal and
  Eq.~\ref{mcothin} therefore likely to underestimate the true disk
  mass \citep[also, e.g.][]{matra2015}.

\paragraph{Estimate of mass limits imposed by CO photodissociation} As noted in previous work \citep{Kospal13,moor2015,kral2017}, survival of CO against photodissociation could  impose stricter constraints on the mass needed to
explain the observed line fluxes, especially for the secondary origin
scenario. For an ambient interstellar field
\citep[e.g.,][]{habing1968}, the lifetime of a CO molecule is $\sim
120$ years \citep{visser2009}.  A rough order of magnitude estimate of the column
densities available for shielding in the two origin scenarios with
masses for emission in LTE as described above yield $N_{H_2} \sim
10^{18}$ cm$^{-2}$ (primordial, $n_{H_2}\sim 10^4$ cm$^{-3}$) and
$N_{CO} \sim 10^{16}$ cm$^{-2}$ (cometary, $n_{CO}\sim 100$ cm$^{-3}$
); at these column densities UV shielding of CO by H$_2$ and CO
self-shielding are not very efficient \citep{visser2009}. Primordial
gas disk masses need to be $\sim 10^4$ higher, $\sim 25$M$_{\oplus}$,
for CO to be shielded and survive for the age of the system.  For the
secondary origin scenario, the CO is only marginally self-shielded and
the estimated CO mass yields lifetimes $\sim 2500$ years; but since
there is continuous CO production, the mass in this case depends on
the replenishment rate with lower disk masses requiring higher
replenishment rates. 

To summarize, we find that simple estimates using
spatial information from the resolved emission maps and typical CO
line luminosities yield plausible masses of a few tens of a
M$_{\oplus}$ of H$_2$ (and $\sim$ few $10^{-3}$ M$_{\oplus}$ of CO)
for a primordial gas disk and $\gtrsim 10^{-3}$ M$_{\oplus}$ of CO
(depending on the CO production rate) for cometary gas to explain the
observed emission.  While the above estimates are informative, the
emission is quite sensitive to the various simplifying assumptions
made; the disk temperature structure and CO photo-chemistry need to be
solved for a more accurate determination of the disk mass and to infer
implications for the origin scenarios. Moreover, CO is likely 
optically thick at these estimated densities affecting the interpreted
mass.

We next use thermochemical models (see Appendix B) that solve for gas temperature and chemistry,
consider gas line emission with non-LTE radiative transfer and fit the observed \coj \ line
emission to calculate disk masses.

\begin{figure*}
\epsscale{0.41}
\hspace{-0.8cm}\plotone{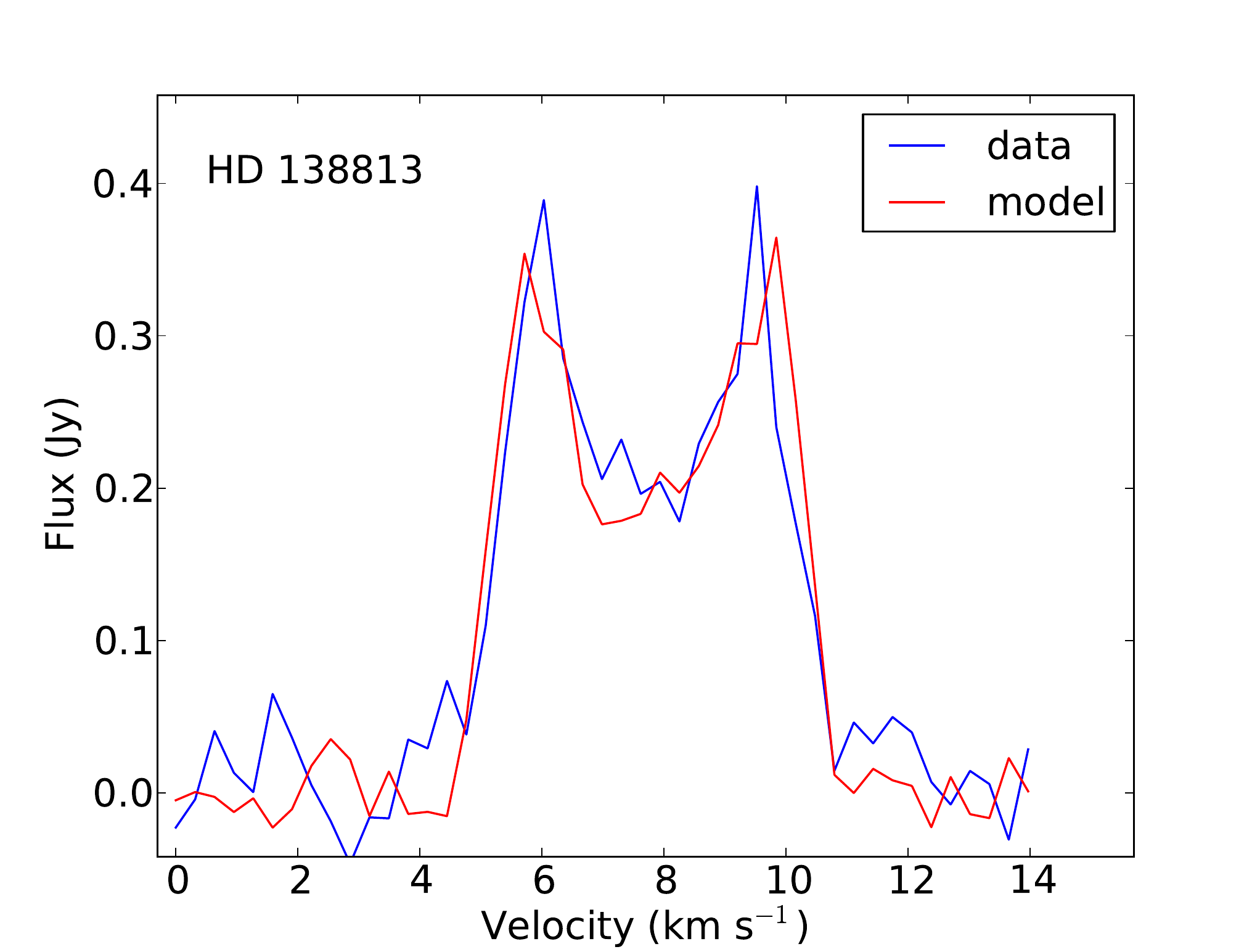}\hspace{-0.8cm}
\plotone{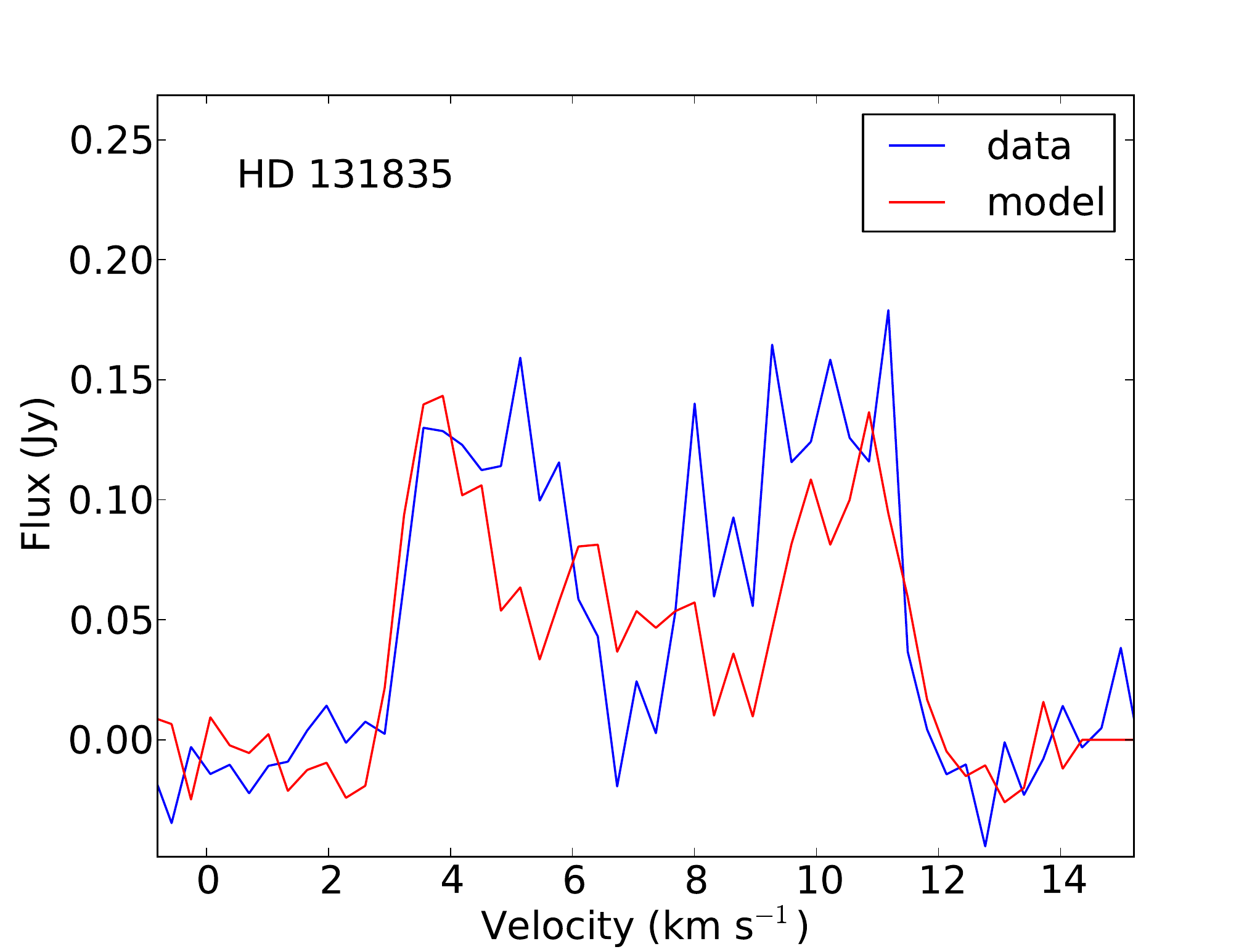}\hspace{-0.6cm}\plotone{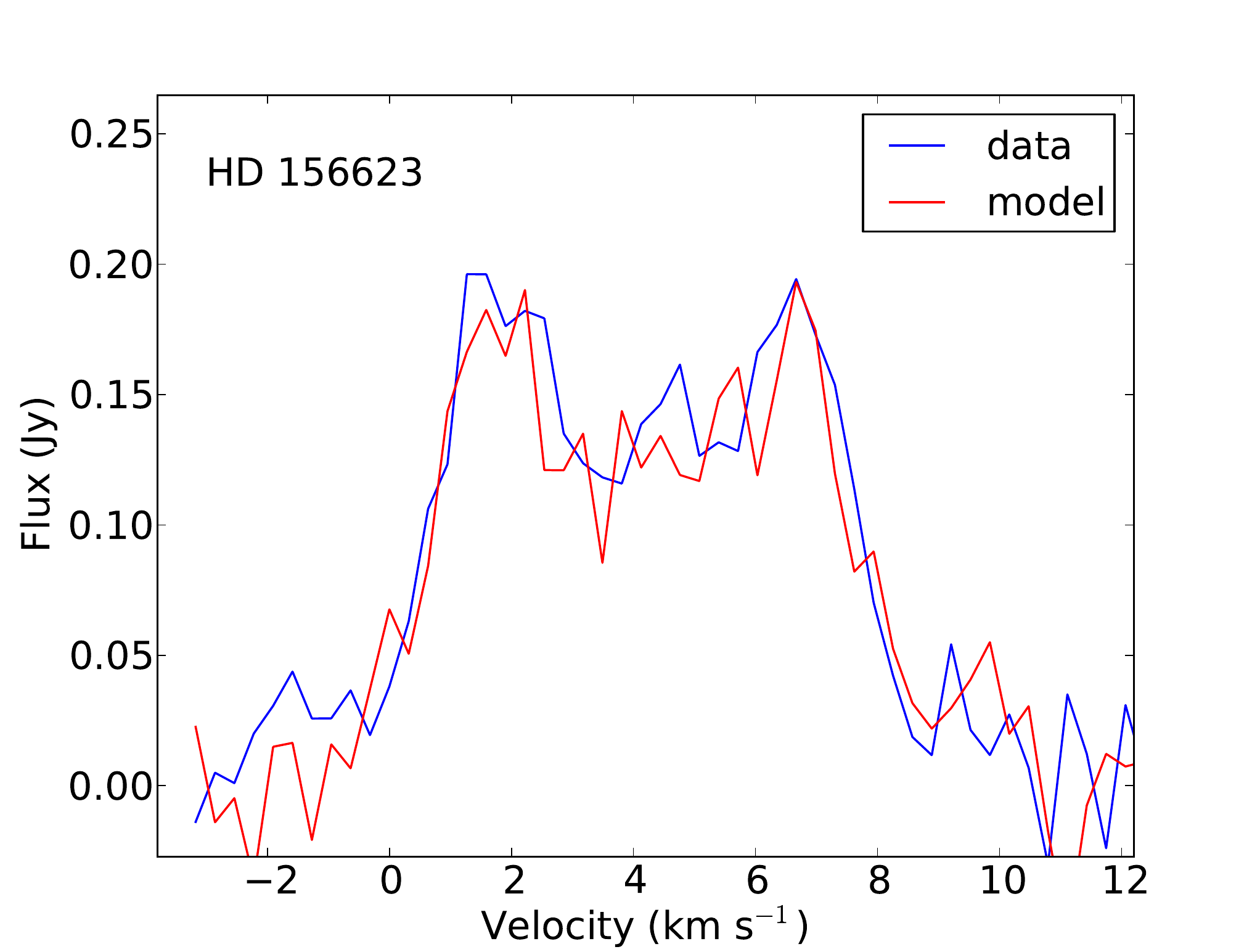}

\caption{Primordial origin models for HD~138813, HD~131835 and
  HD~156623. Spectra were computed by integrating the line images over
  a 2$\arcsec$ aperture. The red lines show the spectra for the
  primordial gas disk models, after processing them through {\sc
      simobserve} to create synthetic visibilities with thermal noise
    added (see Section~\ref{primordial})
}
\label{intspec-pa}
\end{figure*}

\subsection{Modeling gas of primordial origin}\label{primordial}
In these models, we assume typical values for all input parameters and
only vary the gas mass to match the observed CO emission.  We set
initial disk elemental abundances as in the interstellar
medium\citep{jenkins2009}. These values are typical of protoplanetary
disk gas with the elemental C abundance relative to H equal to
$1.4\times10^{-4}$. The dust mass, grain size and radial extent are
from the SED modeling of \citet{Lieman2016}. The dust distribution is
kept fixed. We adopt double power laws to describe the surface
density distribution of gas,
\begin{equation}
\Sigma(r)=\begin{cases}
\Sigma_0 \big( \frac{r}{R_{0}}\big)^{p_{1}} \quad \phantom{\infty}\text{for}\,\, R_{min}<r<R_0   \\
\Sigma_0 \big( \frac{r}{R_{0}}\big)^{p_{2}} \quad \phantom{\infty}\text{for}\,\, R_{0}<r<R_{max} \\
      \end{cases}
      \label{eqn14}
\end{equation} 
where $\Sigma_0 $ is the surface density at $R_0$. We vary the
surface density distribution of gas, and set the local gas/dust mass
ratio accordingly. From the Lucy-Richardson modeling, the inclination
and radial extent of the gas disk is also constrained and the only
free parameters are the radial dependence of the surface density
profile and the integrated gas mass of the disk. We note that while
the Lucy-Richardson modeling gives the emissivity profile, this does
not directly correspond to a surface density, especially when the CO
emission becomes optically thick. Gas heating, cooling and chemistry
that result from the adopted surface density distribution are all then
calculated by the models (for model details see Appendix B). The
surface density profile ($\Sigma_0,R_0,p_1,p_2$) is then varied  as described below until the
synthetic model line emission profile matches the observed line
emission.

 %
 %
 %
We initially fit the integrated flux by varying the total disk mass in
the range 0.1 - 35M$_{\oplus}$ in steps of 5M$_{\oplus}$, to narrow
down the mass range. We then fit the spatially integrated velocity
line profile by varying the disk mass in increments of
0.1M$_{\oplus}$, and then select the best fit models for a more
detailed analysis as follows. The surface density exponents p$_1$
  and p$_2$ were varied manually between -5 and +5 using different
  step sizes, and then fine-tuned using steps of 0.05.
The model data is used to generate synthetic CO line emission data
cubes using LIME \citep{brinch2010}, with non-LTE radiative
  transfer and considering o-H$_2$, p-H$_2$, H and $e^-$ as collision
  partners \citep[][respectively]{Dickinson1977,yang2010,walker2015}. The synthetic
  data cubes are then  processed through
CASA{\footnote{\url{http://casa.nrao.edu/}}} version 5.1.0
\citep{2007ASPC..376..127M} to compare emission in each velocity
channel. The model images were processed through the CASA task {\sc
  simobserve} to create synthetic visibilities, using the same
integration time, spectral setup and antenna configuration used for
the observations as well as injecting the appropriate amount of
thermal noise. The resulting model visibilities were then imaged using
the same {\sc clean} parameters used for the real data.

\begin{figure}
\epsscale{0.9}
\plotone{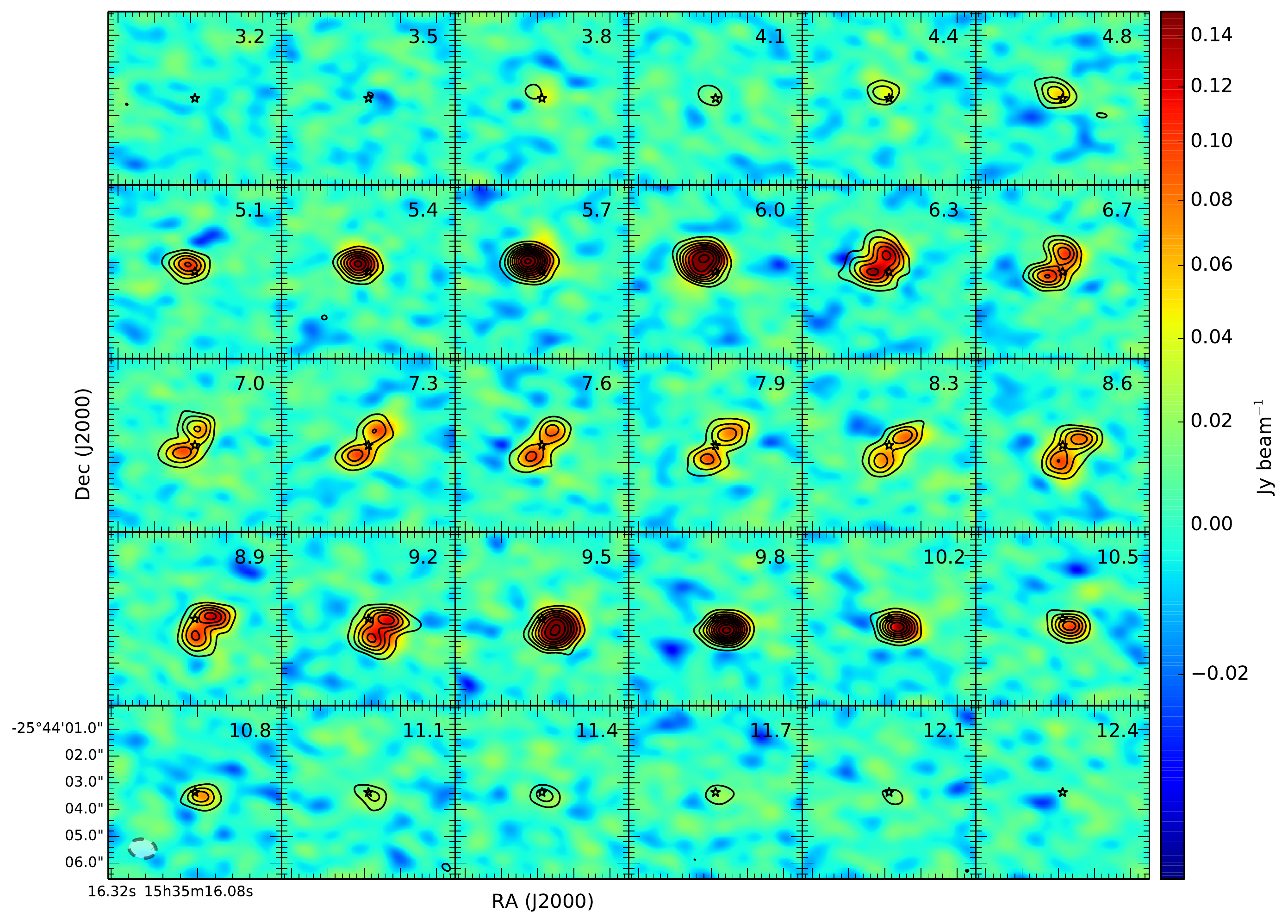}

\caption{Primordial origin model for HD~138813 compared to the
  $^{12}$CO(2--1) ALMA data \citep{Lieman2016}. $^{12}$CO channel maps
  towards HD~138813 (colorscale) with synthetic primordial origin
  model overlaid (contours). Contour levels start at 3$\sigma$ with
  3$\sigma$ intervals (where $\sigma$ is the image rms of 7
  ~mJy\,beam$^{-1}$). The star symbol in the center of each panel
  represents the stellar position.  The velocity of the channels is
  shown in the Local Standard of Rest (LSR) frame, centered at the
  rest frequency of $^{12}$CO(2--1).  }
\label{chanmap-pa-76310}
\end{figure}

The integrated spectra for the best fit models are compared to the
$^{12}$CO(2--1) data in Figure~\ref{intspec-pa}. We obtain total H$_2$
disk masses of 8.1, 7.2 and 10.4 M$_{\oplus}$, and total CO masses of
$5.0\times 10^{-3}$,$4.2\times 10^{-3}$ and $3.8\times 10^{-3}$
M$_{\oplus}$ for HD~131835, HD~138813 and HD~156623 respectively. The
simulated channel maps for the primordial origin model for HD~138813
are compared to the data in Figure~\ref{chanmap-pa-76310}.

\begin{deluxetable}{cccccccc}
\tablecaption{Surface density parameters of primordial and secondary models 
}
\tablehead{
   \colhead{Star} &
   \colhead{Model} &
   \colhead{R$_{0}$} &
   \colhead{$\Sigma_0$} &
   \colhead{R$_{min}$} &
   \colhead{R$_{max}$} &
   \colhead{p$_{1}$} &
   \colhead{p$_{2}$} \\
      \colhead{} &
   \colhead{} &
   \colhead{(au)} &
   \colhead{(g/cm$^2$)} &
   \colhead{(au)} &
   \colhead{(au)} &
   \colhead{} &
   \colhead{} 
}
\startdata
\hline    
HD~131835 &  Primordial  &  80  &    9.8$\times10^{4}$  &   50 &  250  &   -1.5  &  -2.5 \\
  &  Secondary  &  90  &    1.1$\times10^{5}$   &   50 &  250  &   -1.0  &  -4.0 \\
          &        &      &            &      &       &         &       \\
HD~138813 &  Primordial   &  65  &    7.5$\times 10^{4}$   &    5 &  210  &   -1.5  &  -2.7 \\
 &  Secondary  &  65  &    1.0$\times 10^{5}$   &    5 &  210  &   -0.5  &  -4.0 \\
          &        &      &            &      &       &         &       \\
HD~156623 &  Primordial  &  10  &    2.5$\times 10^{3}$   &    5 &  160  &   -0.5  &  -0.5 \\
 &  Secondary  &  10  &    9.2$\times 10^{6}$   &    5 &  160  &   -0.5  &  -0.5 \\
\enddata
\tablecomments {Surface density distributions correspond to double power laws; $\Sigma(r)=\Sigma_0 (r/R_0)^{p_1}$ for $R_{min}<r<R_0$
and $\Sigma(r)=\Sigma_0 (r/R_0)^{p_2}$ for $R_{0}<r<R_{max}$. Here $\Sigma(r)$ refers to the total mass surface density of all species. }
\label{plaws}
\end{deluxetable}

Heating, cooling and photochemistry in the disk around HD~138813 are
described in more detail below; the other two disks are similar in
their chemical and physical structure and not discussed. Heating
mechanisms considered include collisions with dust, X-rays and cosmic
rays, UV grain heating, H$_2$ vibrational heating, H$_2$ formation
heating, exothermic chemical reactions and photo-reactions such as the
ionization of carbon; cooling is by dust collisions, and several
ionic, atomic and molecular lines. Since disk mass is being inferred
using only the CO J=2-1 line, we need only concern ourselves with
heating in the regions where CO resides. None of the three stars have
any detected X-rays, we therefore assumed an X-ray luminosity of
$10^{27}$ erg s$^{-1}$, typical of A stars \citep[e.g
][]{Feigelson2011}. At this level, X-rays do not contribute
significantly to the heating. Dust collisions are important in the
regions with dust (73-161~au for HD~138813). At typical model gas
densities, drag may be sufficient to retain dust grains smaller
than the blow-out size (see Appendix C), but we do not
consider any small dust population that can be retained to heat the
gas \citep[the blow-out size for each disk were taken
  from][]{Lieman2016}.  We assume that a small fraction ($<$1\%)
  of the grains are spatially co-located with the gas outside the
main dust belt (this may be possible because of gas drag). We note
that such small amounts of dust are still consistent with the SED
fitting used to determine dust parameters.  Photoelectric grain
heating (due to FUV photons) in the absence of very small grains and
Polycyclic Aromatic Hydrocarbons (PAHs) is not very significant, and
here dust heating is only due to relatively large $\sim 1\mu$m grains
\citep{Kamp2001}. The other main heating mechanisms are UV pumping,
  H$_2$ formation and cosmic rays (we use $\zeta_{CR}\sim
10^{-16}$ s$^{-1}$). The cosmic ray rate in the Galaxy could be as
  high as $10^{-15}$ s$^{-1}$ \citep[e.g.][]{Indriolo2012,
    Neufeld2017}. If we assume $\zeta_{CR} \gtrsim 10^{-16}$ s$^{-1}$,
  then cosmic ray heating typically dominates gas heating. In this
  case, cosmic rays provide the necessary heating (to excite the CO
  line) and there is no need to assume that there is any small dust
  that gets dragged along with the gas.  Figure~\ref{heat} shows the
dominant heating in the disk multiplied by the CO number density to
emphasize the regions where CO is present. The main coolants in the
disk are CO rotational emission and [OI] fine structure emission.
The resulting temperature and density structure is shown in
Figure~\ref{disksDensTemptructure}. The gas surface density parameters 
of the best fit models are presented in Table~\ref{plaws}.

\begin{figure}
\centering
\includegraphics[scale=0.9]{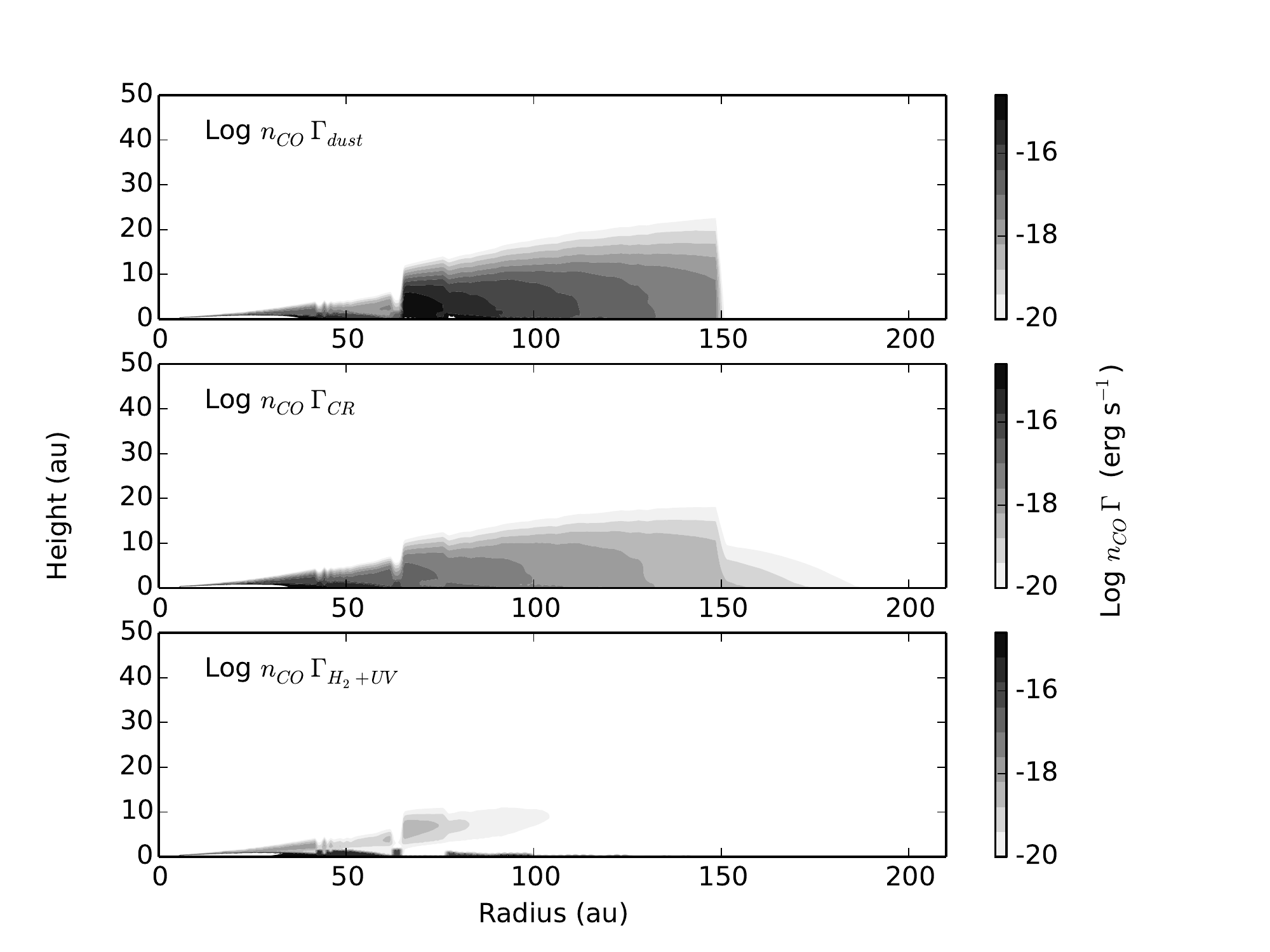}
\caption{The main heating mechanisms in the disk around HD~138813 for the primordial case,
  with the heating rate multiplied by the CO number
  density. $\Gamma_{dust}$ includes heating by dust collisions in the
  belt and grain photoelectric heating outside this region. Heating by
  cosmic rays ($\Gamma_{CR}$) and H$_2$ vibrational heating and H$_2$
  formation heating are shown in the middle and lower panels. The
  latter can be the most dominant heating mechanism in the mid plane
  of the disk. }
\label{heat}
\end{figure}

\begin{figure}
\centering
\plottwo{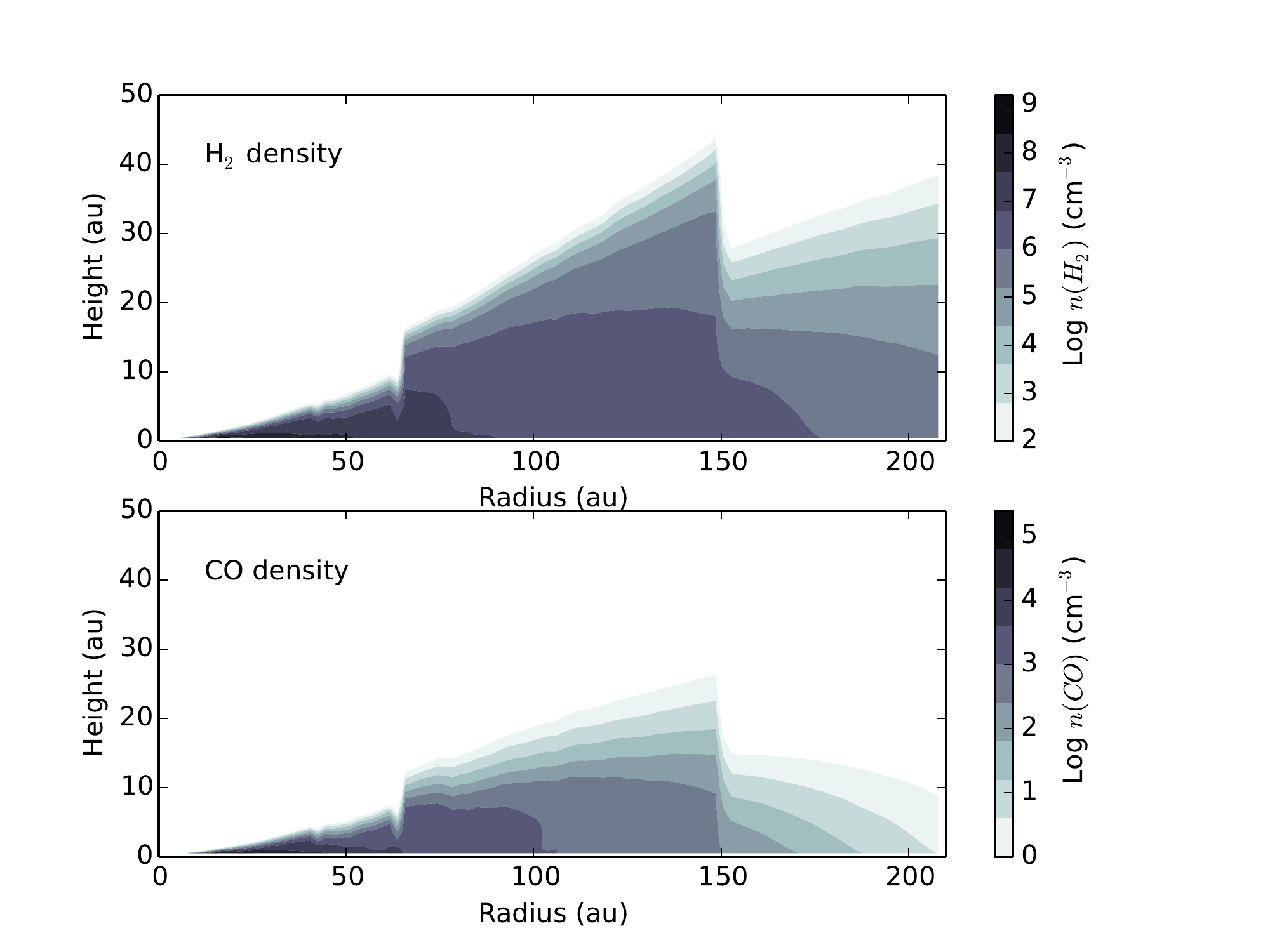}{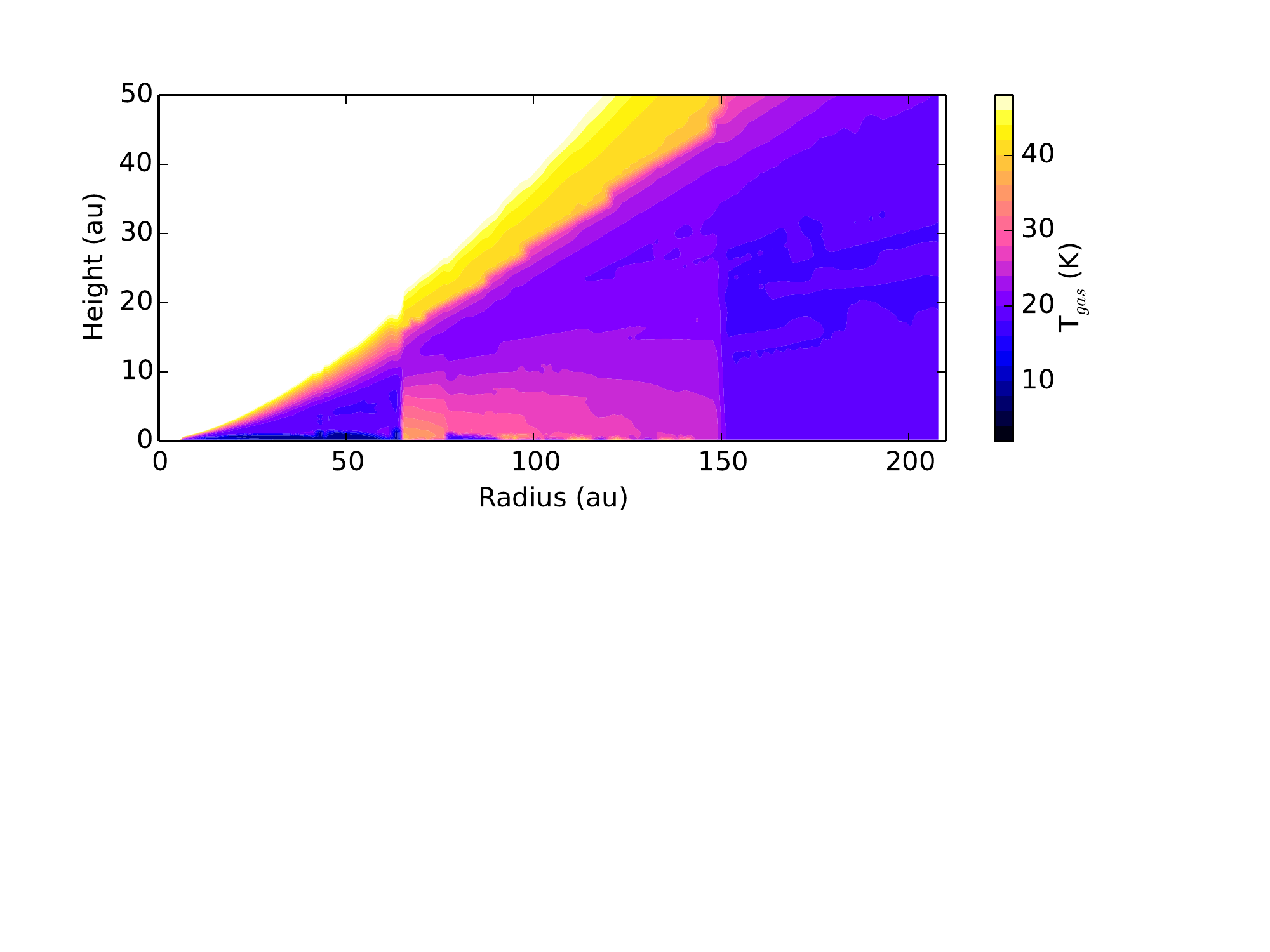}
\caption{Temperature and density structure for HD~138813 for the primordial case. The CO density closely follows the H$_2$ density in the
disk, and is higher in the dust belt where there is more efficient
formation of H$_2$. Dust collisions raise the temperature in this
region of the disk. The disk gas outside this region is in general
$\lesssim 20$ K.}
\label{disksDensTemptructure}
\end{figure}

Heating by dust in the $67-147$~au
region results in an increase in the temperature and therefore higher
pressure leading to a more vertically extended disk. Outside the dust
belt, the gas temperature is quite low, and in general insufficient to
excite the J=2--1 line of CO in the outer disk.

Since the primordial origin disk is optically thick in CO, the
resulting emission is relatively insensitive to mass and depends on
the disk temperature structure. Uncertainties in disk heating
mechanisms translate to the calculated gas temperature, and could
potentially result in different disk mass estimates for the same line
emission flux.  However, the disk mass also affects the chemistry and
the amount of CO in the disk.  H$_2$, CH, C and CO have significant
cross-sections in the 11.3-13.6eV energy range and absorb incident UV
flux to shield and preserve CO deeper in the disk, with H$_2$
shielding being the most effective. As discussed earlier, the column
densities required are such that the implied masses for CO to survive
are high.  An additional complication is the paucity of dust in the
disk. We assumed that there was dust at the 1\% level distributed
through the disk; not only does this provide some additional heating
to raise the gas temperature outside the dust belt, but also enables
formation of molecular hydrogen that can shield CO. Increasing this
dust fraction affects the SED-fitting, while lowering it increases the
required mass due to inefficient H$_2$ formation (and
CO-shielding). As discussed previously, increasing the cosmic ray
ionization rate could also lead to more gas heating.  Therefore,
although we could drive the disk to lower masses with an increase in
gas temperature especially in the outer disk and still reproduce the
observed emission, the limiting mass necessary for self-shielding
makes it very unlikely that the primordial disk mass can be
significantly lower than calculated from the models.


Disk masses higher than derived from the modeling are not ruled out. Decreasing the heating in the disk, for example, by lowering the cosmic ray rate and/or reducing the dust density, lowers the gas temperature (note that increasing the gas mass reduces the effective  cross-section of dust per H nucleus and decreases the gas  temperature even as the dust mass is unchanged). Lower temperatures in a more massive disk are consistent with the observed emission as well, and this degeneracy cannot be broken with only one CO transition observed.\footnote{Non-detections of [OI] and [CII] with Herschel PACS  \citep{mathews2013, moor2015} are not very useful constraints on disk mass, as the flux upper limits are a factor of $\sim$ 100 or  more higher than calculated fluxes from the best-fit models for  these lines.}  However, disk masses cannot be considerably higher than derived here because then  the densities become high enough for gas drag to prevent radiation pressure from removing grains in the debris collisional cascade, (see Appendix C). Lack of removal of grains  (which is size-independent as both forces depend on grain area) will result in an accumulation of small grains making the dust disk optically thick and inconsistent with the debris disk classification.

\begin{figure}
\epsscale{0.41}
\hspace{-0.8cm}\plotone{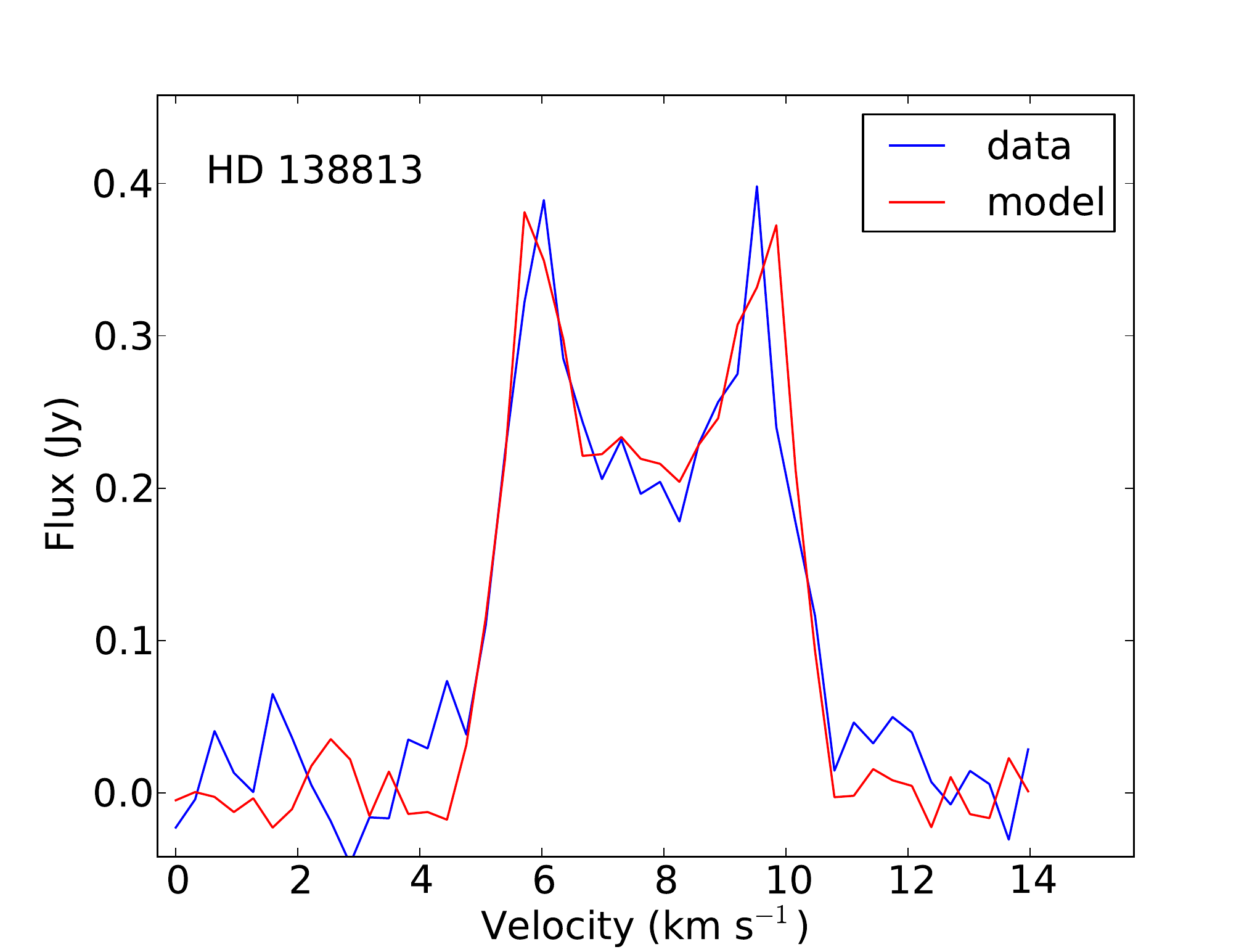}\hspace{-0.8cm}
\plotone{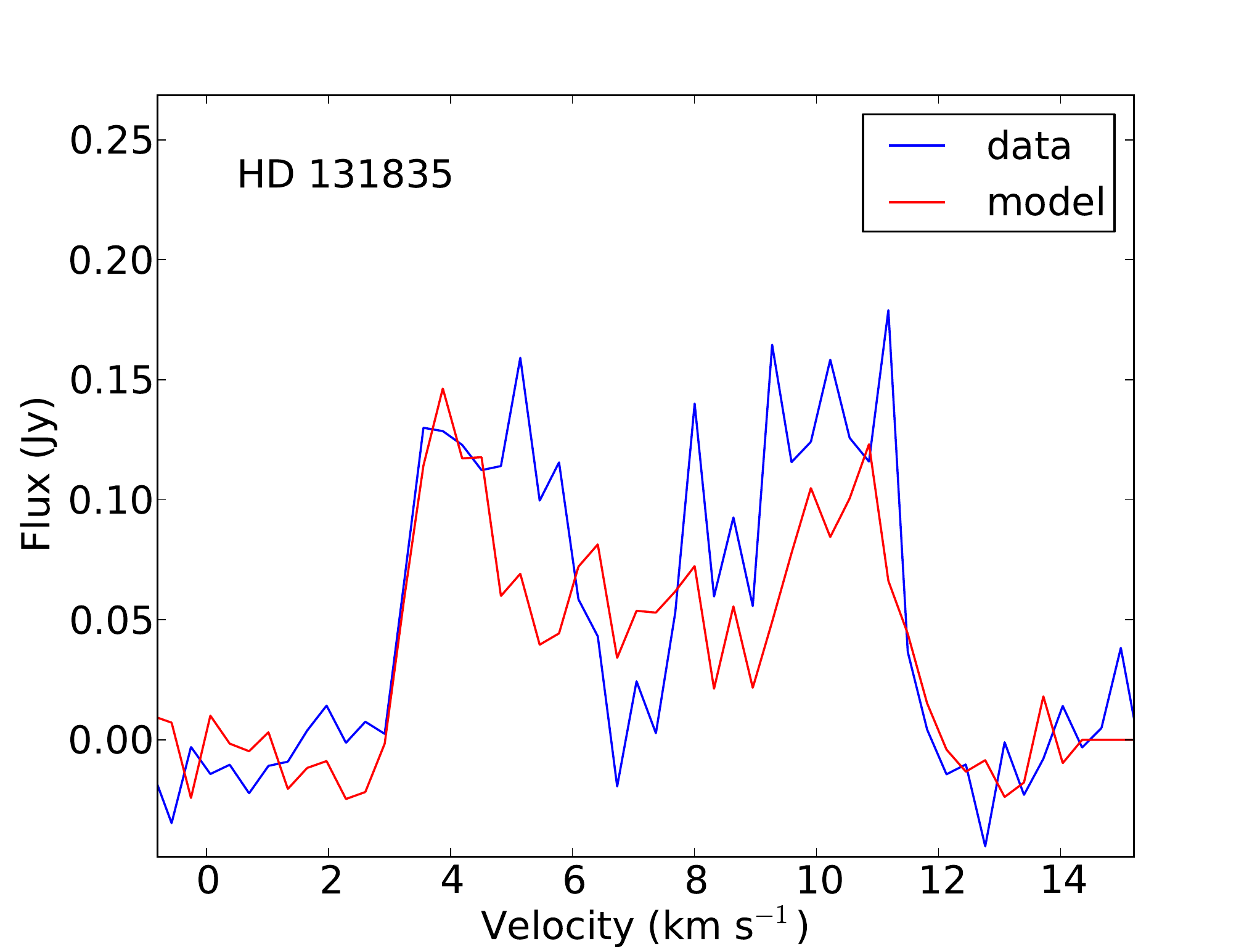}\hspace{-0.6cm}
\plotone{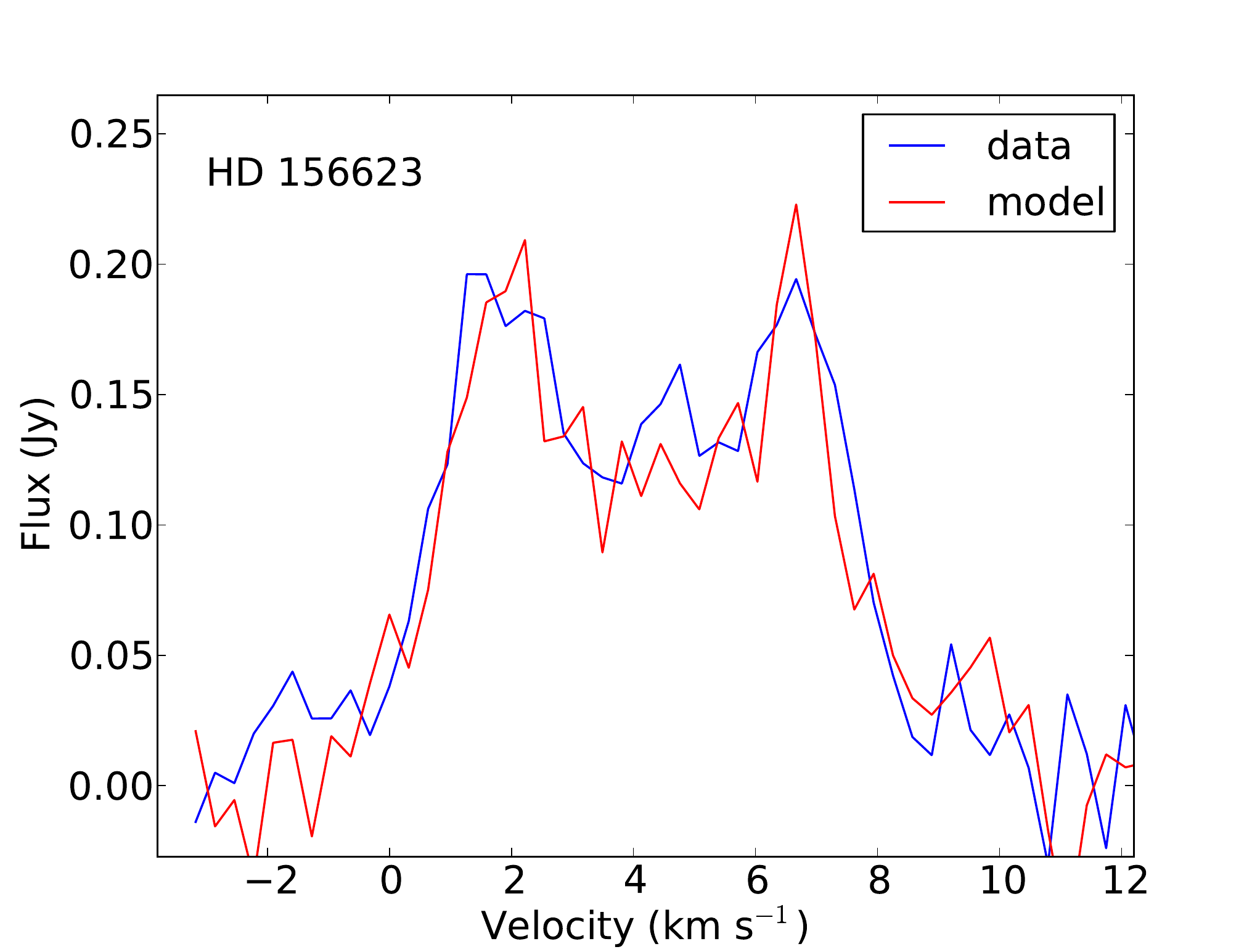}

\caption{Secondary origin models for HD~138813, HD~131835 and HD~156623. 
}
\label{intspec-ca}
\end{figure}

\subsection{Modeling gas of secondary origin}

 In the cometary model, CO may be released by thermal and UV
 desorption of ices from dust grains \citep{grigorieva2007}, or by
 high velocity impacts of larger bodies \citep{zuckerman2012}. In the
 latter case, the heat generated by the impact would release CO into
 the gas phase, and the initial temperature of the CO gas would be set
 by the energetics of the collision which heat the gas. If CO is
 sublimated from dust, the gas that is released is expected to be
 approximately at the temperature of the dust grain. After desorption,
 the gas will initially expand and cool adiabatically and radiatively
 and will be heated further out in the flow by photo and chemical
 processes \citep[e.g.][]{rodgers2002}. A full and accurate treatment
 of the gas temperature requires a chemo-dynamical model, possibly
 including impact collision modeling, and is beyond the scope of this
 work.  Here for simplicity, we assume that the gas temperature is
 equal to the local dust temperature. However, we fully solve for the
 chemical evolution of the gas; we assume that at the start of the
 simulation CO and H$_2$O are released in a 1:10 ratio
 \citep{mumma2011}, and our initial abundances are such that all the
 gas is in CO and H$_2$O. We consider different initial surface
 density distributions of the gas and solve for time-dependent
 chemical evolution of the disk after the CO and H$_2$O are
 released. We use the same chemical network as described in Appendix
 B, but now only solve for the dust temperature and set the gas
 temperature equal to it.  For low initial surface densities,
 molecules are rapidly photodissociated unless there is a sufficient
 self-shielding. The disk surface density at $t=0$ is increased to a
 point where a column of CO is built up that can explain the observed
 CO emission. We typically run the chemical model up to times of ~2
 Myr. The best fit cometary gas masses are $4.6 \times 10^{-3},
 3.1\times 10^{-3}$ and $2.45\times 10^{-3}$ M$_{\oplus}$ for CO alone
 (with total gas content being $\sim 15$ higher) for the disks around
 HD~131835, HD~138813 and HD~156623. The CO masses derived are higher
 than the minimum mass optically thin estimate, and almost meet the
 mass needed for LTE.  The temperature of the gas is higher than $\sim
 20$K through most of the disk, and in the outer disk, densities are
 low enough that n-LTE conditions (collisions with neutrals and
 electrons) result in sub-thermal emission.  The gas surface
   density parameters of the best fit models are presented in
   Table~\ref{plaws}.
 

Figure~\ref{intspec-ca} compares the integrated spectra for best fit
models of cometary origin to the $^{12}$CO(2--1) data
\citep{Lieman2016}.  The models have been used to generate synthetic
spectra using LIME, and then through CASA to produce simulated
visibilities and channel map images comparable to the $^{12}$CO(2--1)
ALMA data. The simulated channel maps from the secondary origin model
for HD~138813 are compared to the data in
Figure~\ref{chanmap-ca-76310}.

\begin{figure}
\epsscale{0.9}
\plotone{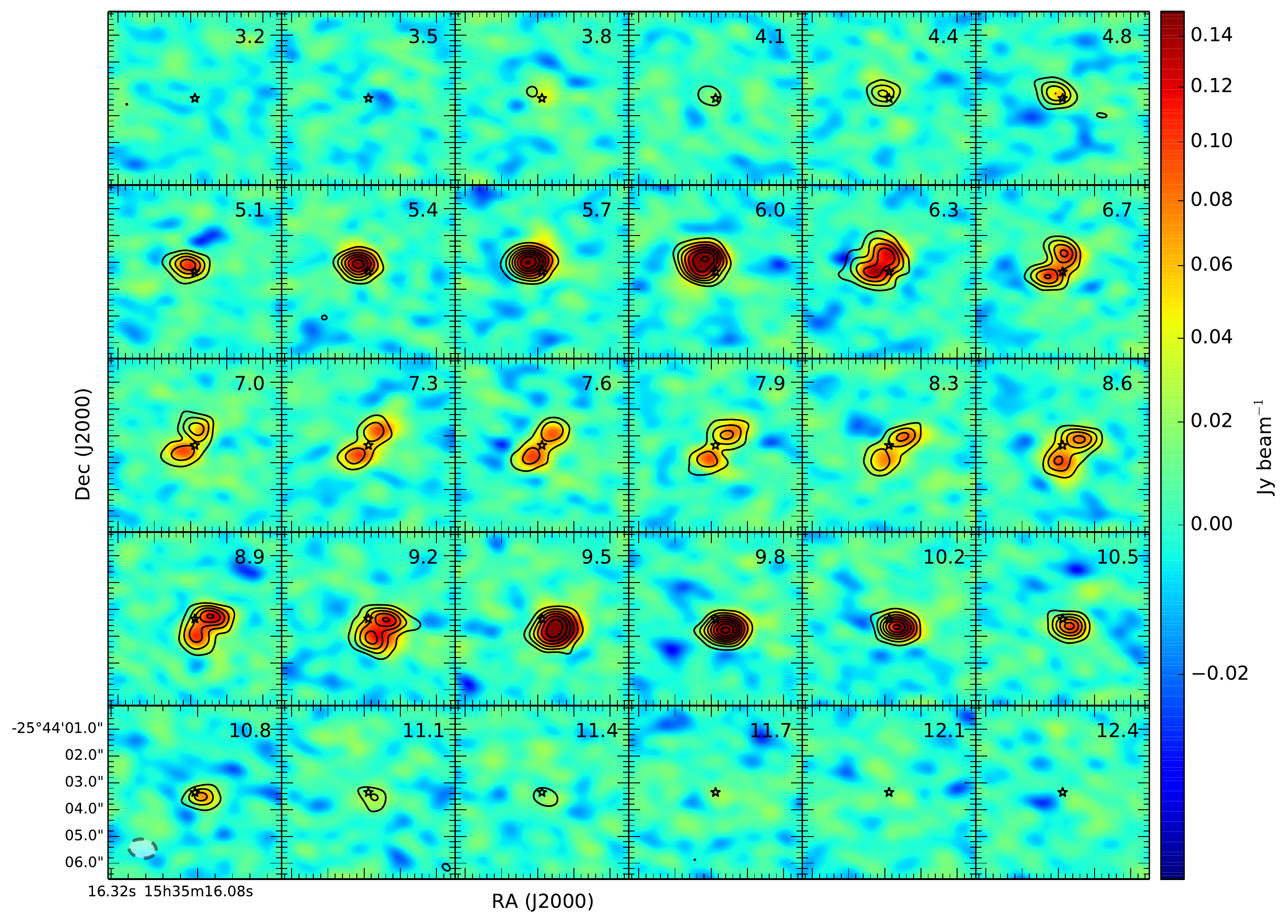}

\caption{Secondary origin models for HD~138813 compared to the
  $^{12}$CO(2--1) ALMA data \citep{Lieman2016}. $^{12}$CO channel maps
  towards HD~138813 (colorscale) with synthetic cometary origin
  model overlaid (contours). Contour levels start at 3$\sigma$ with
  3$\sigma$ intervals (where $\sigma$ is the image rms of 7
  ~mJy\,beam$^{-1}$). The star symbol in the center of each panel
  represents the stellar position.  The velocity of the channels is
  shown in the Local Standard of Rest (LSR) frame, centered at the
  rest frequency of $^{12}$CO(2--1). .
}
\label{chanmap-ca-76310}
\end{figure}


As pointed out by previous studies
\citep[e.g][]{zuckerman2012,Kospal13,Dent2014,kral2017}, simple
estimates suggest that CO survival against photodissociation is
limited to $\sim 120$ years; a continuous replenishment of CO by
cometary collisions or outgassing is therefore needed to explain the
observed emission.  For the best fit disk masses, the CO column
densities are high enough ($10^{16-17}$ cm$^{-2}$ at $\sim 80$au) for
self-shielding of CO and shielding by other C-species to reduce the
photodissocation rate by $\sim$ 2 orders of magnitude. The CO
photo-dissociation rate in the disk is moreover not constant, and can
decrease in time due to an increase in the gas opacity \citep[due to many
C-species, and primarily neutral carbon; also see][]{kral2018} built up in the disk as CO is
destroyed with time (which however also reduces self-shielding), and
this can be calculated from the chemical disk model. CO is also
re-formed in the gas phase due to the abundance of O-species (from the
destruction of H$_2$O) which further lowers the net CO destruction
rate.  In order to maintain a column of gas that can shield CO against
UV photons, the production rate needs to be equal to the net
destruction rate. We estimate this by considering the initial mass of
the disk (at $t=0$), the timescale $t_f$ taken to reach the final mass
of CO needed to explain the emission, and therefore the production
rate $\sim (M(t=0) - M(t_f))/t_f$. Models with higher initial disk
masses lead to higher $t_f$ as is to be expected, and give
approximately the same production rates.  We note that disks with
total gas masses (including water released by collisions) of $\sim 4
\times 10^{-2}$ \mearth\ are needed to obtain CO masses $\sim 3\times
10^{-3}$ as derived here, and imply a replenishment rate of $\sim
1.6\times 10^{14}$ g s$^{-1}$ ($8\times 10^{-7}$
  M$_{\oplus}$~yr$^{-1}$) to retain the minimum mass of CO needed. Given
that to capture the true evolution of the secondary gas disk, one
needs to consider chemistry with hydrodynamical flow, we do not
attempt any more detailed estimates of the production rate.

\section{Discussion}\label{discussion}

Based on the above modeling of the CO J=2-1 line alone, we find that
both the primordial and cometary disk scenarios are seemingly
consistent with the ALMA data.

Primordial gas indicates the survival of small amounts of H$_2$ gas
long past the main planet formation epochs. While current theories of
disk evolution \citep[e.g.][]{Alexander2014,Gorti2015} predict a very
rapid inside-out dispersal of the entire disk when the disk masses
become too low, it is nevertheless possible for disk lifetimes to
exceed the $\sim 2\times10^7$ year system ages if the disks have a
very low viscosity and low mass loss rates.  If disk evolution
proceeds such that by a few Myr, most of the dust has been
incorporated into larger bodies including planets, leaving $10^{-2}$
M$_{\oplus}$ masses of dust and nearly 1000 times as much gas, the low
abundances of small dust \citep[and hence lower rates of FUV-driven
  photo evaporation, see][]{Alexander2014,Gorti2015}, and the low
X-ray luminosities of these stars may result in the disk surviving to
the so-called hybrid stage. However, we argue that such a scenario is
unlikely. As shown in Appendix C, for disks with masses $\gtrsim
5$M$_{\oplus}$, gas densities become too high, and collisional
coupling of gas and dust grains due to gas drag becomes important. For
secondary dust generated by collisions, small dust grains in the
cascade are retained and would quickly render the disk optically
thick. These disks would then no longer be classified as debris disks.
All three disks modeled here are estimated to have larger H$_2$ masses
(a few tens of M$_{\oplus}$) in the primordial scenario and are
therefore inconsistent with the hybrid disk picture. A similar
conclusion on the effects of gas drag was recently made
\citep{kral2018} even for disks of secondary origin, and hence
considerably lower mass. Further, a primordial origin for disks} does
not explain why most of the CO-bright debris disks should be
preferentially detected around stars lying within a relatively narrow
range of spectral type and mass \citep[e.g.][]{pericaud2017}.

We prefer the secondary origin scenario for the three disks in this
study, as also argued by \citet{kral2017,kral2018}. The partial
co-location of both gas and dust in 2 out of the 3 targets argues in
favor of a cometary origin for the gas. The gas surface density
distribution we derive for HD~131835 is consistent with that obtained
from MCMC fits to CI data \citep{kral2018}, where most of the emission
is confined to a ring located at $\sim$90~au with a 80~au
width. \citet{kral2018} further constrain the range of allowed masses
by considering the CO mass derived by \citet{moor2017} using
observations of the optically thin C$^{18}$O(3-2) line, which is
dependent on other assumptions in their model on viscous spreading and
gas production scenarios. The CO gas mass that we derive for the
cometary production for HD~131835 is about an order of magnitude
lower, but we note the CI line optically thick and hence more
sensitive to temperature than mass. It is likely that the gas
temperature is closer to the CO thermal desorption temperature of
$\sim 20$K and not as high as the dust temperature as we have assumed;
in this case the inferred CO masses can be higher.  In future work, we
plan a multi-line analysis that includes isotope chemistry and a
thermal balance calculation to better constrain the mass of the
cometary gas disk.


For the CO mass estimated, we find that the required replenishment
rates are only slightly higher than that for $\beta$ Pic;
\citet{Dent2014} estimate $2.3\times 10^{-7}$ M$_{\oplus}$~yr$^{-1}$
whereas the implied rate of production for these debris disks is about
a factor of a few higher. Note that if the CO gas temperature is lower
than the dust temperature, and instead close to the grain thermal
desorption value of 20K, then the required production rate is
$1.6\times 10^{-7}$ M$_{\oplus}$~yr$^{-1}$ and lower than that for
$\beta$ Pic. Our rates are also in reasonable agreement with
\citet{kral2018}, despite the differences in modeling approach (they
do not explicitly solve for the chemistry or ionization, while we do
not consider viscous spreading or include a production rate in our
models).


\citet{kral2018} propose that accumulation of
atomic carbon over time results in a layer that shields CO, and that
dissociated CO piles up as atomic carbon prolonging the lifetime of CO. Although we do 
consider the effects of molecular gas opacity and self-shielding in our models and, contrary to \citet{kral2018}, further solve the chemical network  to model the CO emission,
we do not consider the viscous time evolution of the disk.
We however also find that solving for disk chemistry results in CO
lifetimes that are long enough to somewhat mitigate the issue of high
cometary collision rates previously inferred.  It is possible that the
true CO production rate required is even lower that what we estimate
because of the following. The photodissociation rate of CO is
sensitive to the column density profile which determines the
self-shielding and gas opacity factors; a clumpy distribution of
outgassing comets is likely to be more effective at shielding CO. Gas
is also likely to entrain smaller dust particles than considered here,
which are more efficient at attenuating UV photons, reducing the
destruction of CO even further.

 The fact that the dust and CO gas disks are only partially co-located
 in the disks may weaken the secondary origin scenario
 (Figure~\ref{diskradii}), unless the CO disk viscously spreads after
 it has been generated \citep{matra2017a}. If disks are indeed
 MRI-active and accrete efficiently \citep[e.g.][]{kral2016}, then the viscous timescales could
 be short enough that the gas disk does not remain co-spatial with the
 dust on release \citep{kral2018}.

Interestingly, the stellar spectral types are such that the UV flux in the energy range
required for desorption \citep[$\sim 8-9$eV,][]{Fayolle2011} is high
relative to the energy range ($11.26-13.6$ eV) required for CO
photodissociation.  For later spectral types, there are not enough
photons to desorb CO while for earlier spectral types there are too
many that dissociate CO.
However, a normal interstellar field (Habing field G$_0=1$, compared to
G$_0=2000$ at 1au for HD~138813) begins to dominate at $r>45$ au and
unless these disks are in an unusually low ambient UV field
($G_0<0.2$), the change in stellar photon flux going from $\sim 8$ eV
to $\sim 11$eV may not be that important. Debris disks are however quite dusty, and scattering of UV photons even at $\sim 1-2$\% efficiency (ignored in this paper) could keep the stellar radiation field dominant out to $\sim 100$ au where the gas is located. While gas around an M dwarf was recently detected by \citet{matra2019} who suggest that higher CO production rates from collisional cascades for more luminous stars coupled with sensitivity issues may be responsible for the CO detection bias, the ultraviolet spectrum may play an additional role. The UV spectra of M dwarfs are often flare-driven and the flare spectra are in fact found to resemble A star spectra \citep{kowalski2013}.

\section{Conclusions}\label{conclusion}

We have applied the Lucy-Richardson deconvolution technique to derive
the $^{12}$CO(2--1) surface brightness distribution around three gas
rich debris disks detected with ALMA. The derived disk parameters in
conjunction with detailed thermochemical models are used to test
whether the observed gas is a primordial fossil from the
protoplanetary disk phase, or produced by cometary collisions. We find
that while both scenarios can reproduce the observed emission, the
primordial scenario is unlikely. We conclude that gas drag may
dominate small dust dynamics at the few tens of M$_{\oplus}$ disk
masses of H$_2$ needed to explain the observed CO emission, and that
this small dust would not keep a primordial disk optically thin in
continuum emission as observed. Several lines of reasoning favor the
secondary origin scenario: in at least 2 of the 3 disks observed, the
gas and dust are spatially co-located, the production rates we derive
are only slightly higher than those derived in disks of clear
secondary origin such as $\beta$ Pic, and the different photon
energies needed for desorption and photodissociation may explain the
higher detection rates of gas around A stars.


\section*{Acknowledgments}

We thank the anonymous referee for reading the paper carefully
  and providing thoughtful comments which improved the quality of this
  publication.  JMC acknowledges support from the National Aeronautics and
Space Administration under Grant No. 15XRP15-20140 issued through the
Exoplanets Research Program. AMH acknowledges support from NSF grant
AST-1412647. UG was supported by NASA grants NNH13ZDA017C (NAI CAN 7)
and NNX14AR91G (Astrophysics Data Analysis Program).  This paper makes
use of the following ALMA data: ADS/JAO.ALMA\#2012.1.00688.S. ALMA is
a partnership of ESO (representing its member states), NSF (USA) and
NINS (Japan), together with NRC (Canada) and NSC and ASIAA (Taiwan),
in cooperation with the Republic of Chile. The Joint ALMA Observatory
is operated by ESO, AUI/NRAO and NAOJ. The National Radio Astronomy
Observatory is a facility of the National Science Foundation operated
under cooperative agreement by Associated Universities, Inc. This work
has made use of data from the European Space Agency (ESA) mission {\it
  Gaia} (\url{https://www.cosmos.esa.int/gaia}), processed by the {\it
  Gaia} Data Processing and Analysis Consortium (DPAC,
\url{https://www.cosmos.esa.int/web/gaia/dpac/consortium}). Funding
for the DPAC has been provided by national institutions, in particular
the institutions participating in the {\it Gaia} Multilateral
Agreement.

\software{Common Astronomy Software Applications \citep{2007ASPC..376..127M}, {\sc LIME} \citep{brinch2010}, Astropy \citep{2013A&A...558A..33A}, EMCEE \citep{foreman2013}}

%
%

%
%
%

\appendix
\section{Lucy-Richardson MCMC results for HD~131835 and HD~156623}\label{lucyappendix}

\begin{figure*}
\epsscale{0.6}
\plotone{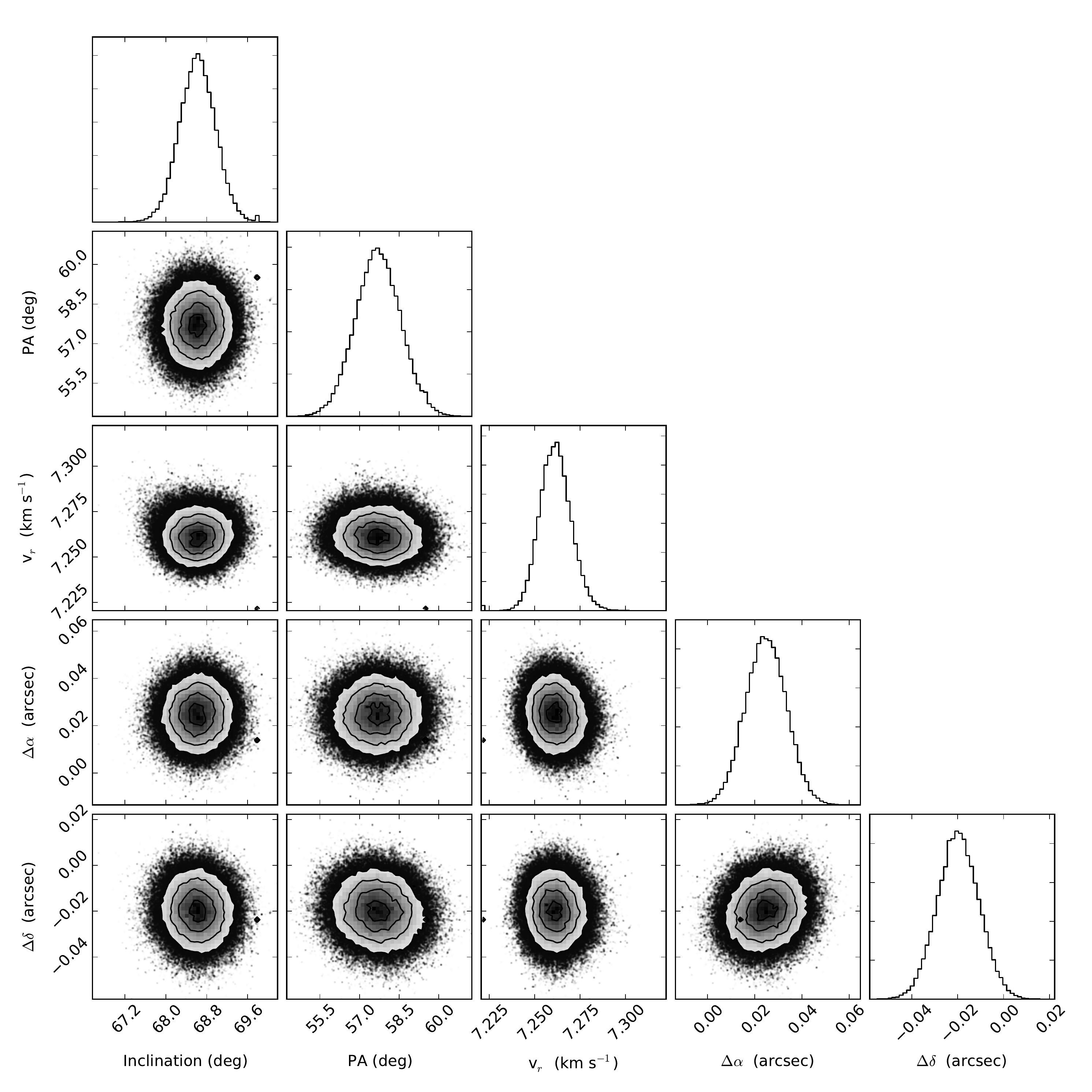}
\caption{ Results of the MCMC chain for HD~131835, showing the one and
  two dimensional projections of the posterior probability. The first
  200 steps corresponding to the burn-out phase are not plotted.  }

\label{fig_corner_73145}
\end{figure*}

\begin{figure}
\epsscale{0.6}
\plotone{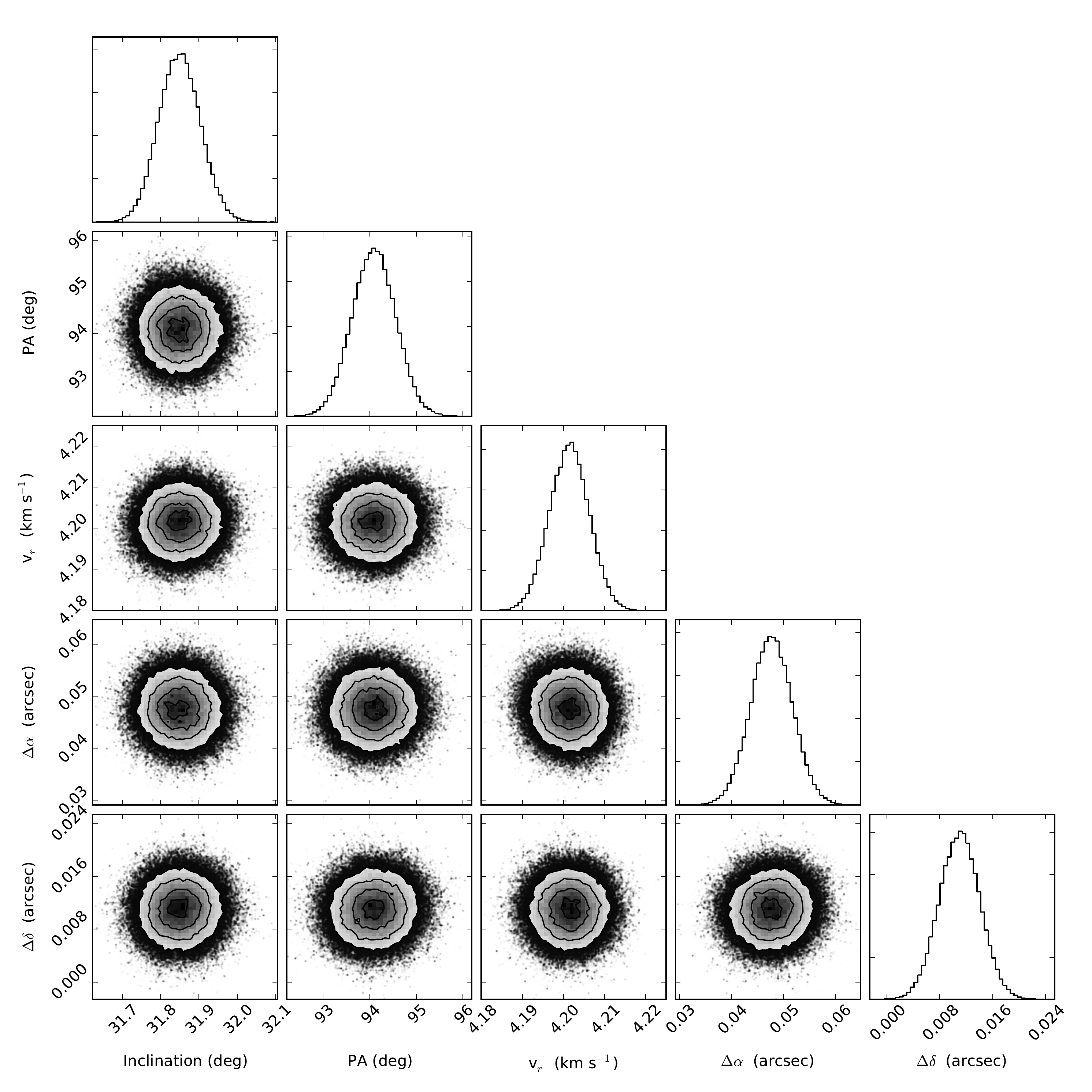}
\caption{Results of the MCMC for HD~156623, showing the one and two
  dimensional projections of the posterior probability. The first 200
  steps corresponding to the burn-out phase are not plotted.  }
\label{fig_corner_84881}
\end{figure}

\begin{figure}
\epsscale{0.9}
\plotone{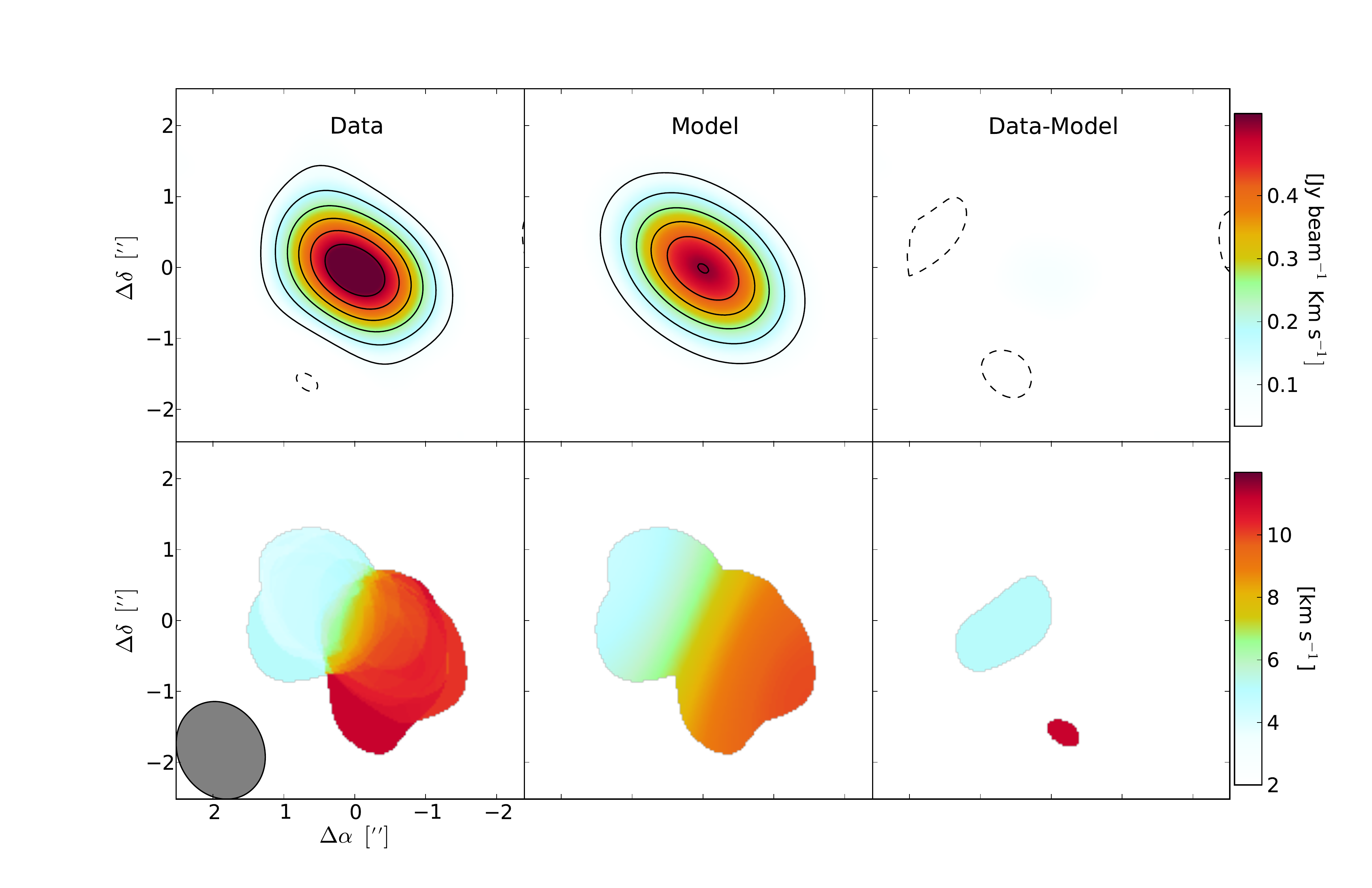}

\caption{
  {\it{Top}}: Integrated intensity map for the data, model and residuals of the  CO emission of the  HD~131835 disk. Contours start at 5$\sigma$ with intervals of 5$\sigma$. Negative contours (dashed lines) start at -2$\sigma$ with intervals of -5$\sigma$.
 {\it{Bottom }}: Intensity-weighted mean velocity (moment 1) for the data, model and residuals of the  CO emission of the HD~131835 disk. 
}
\label{fig:hd73145-residual}
\end{figure}

\begin{figure}
\epsscale{0.9}
\plotone{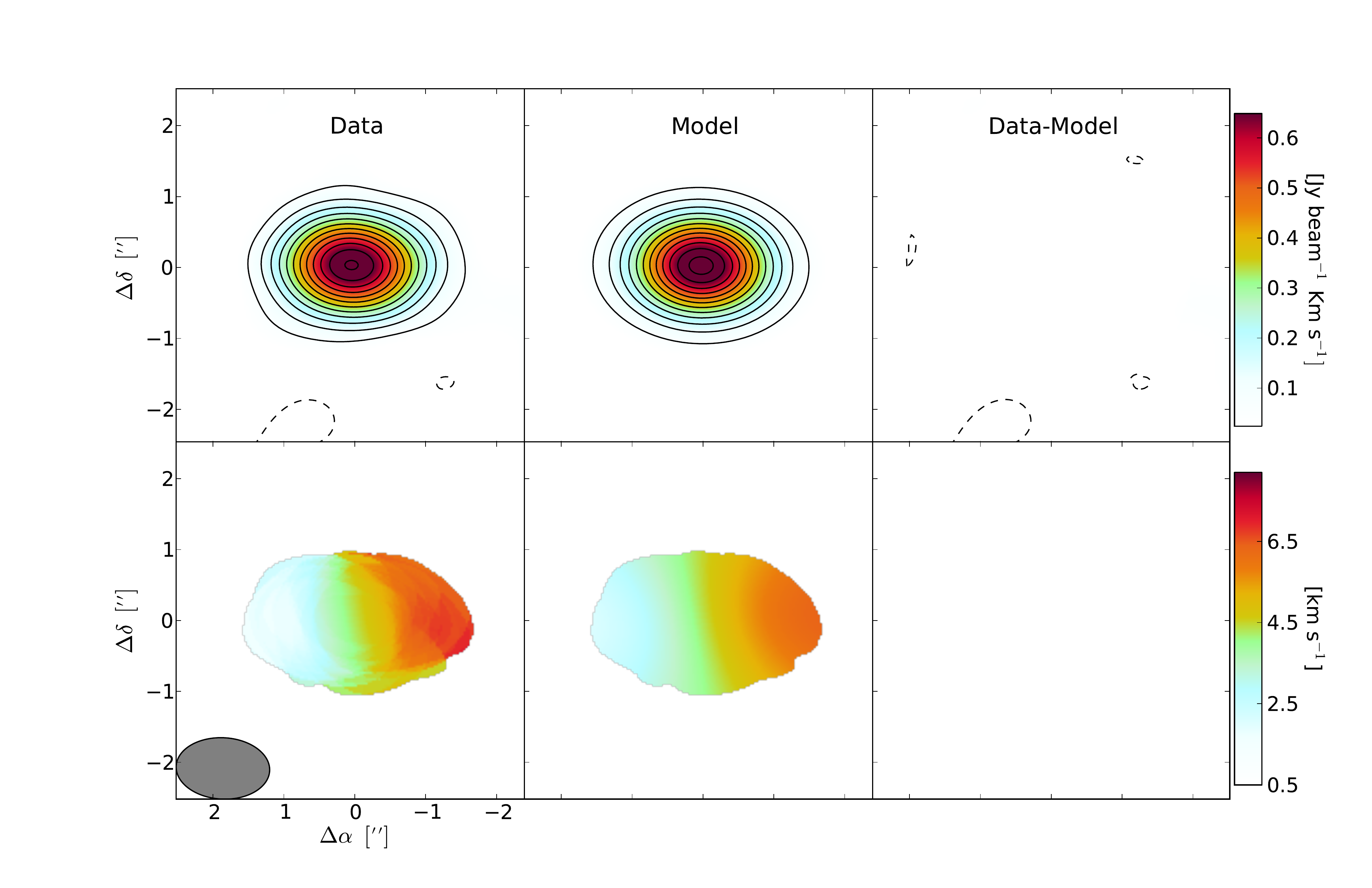}
\caption
    {
      Same as Figure~\ref{fig:hd73145-residual} but for the HD~156623 disk.%
}
\label{fig:hd84881-residual}
\end{figure}

Figures~\ref{fig_corner_73145} and ~\ref{fig_corner_84881} show the
MCMC results from the Lucy-Richardson deconvolution for HD~131835 and
HD~156623 (Section
~\ref{modelling}). Figures~\ref{fig:hd73145-residual} and
~\ref{fig:hd84881-residual} show the comparison between the
corresponding best-fit models and the ALMA data.

\section{Thermochemical model details}
The disk models used here were first developed for modeling gas in debris disks \citep{Gorti2004} and later extended to apply to all gaseous protoplanetary disks in general \citep{Gorti2008, Hollenbach2009, Gorti2011, Gorti2015}. We briefly summarize the main aspects of the models here and refer the reader to the original papers for more details. 

The models assume azimuthal symmetry and solve for the density and temperature structure and chemistry in a self-consistent iterative manner, and calculate the gas and dust temperatures separately. They include heating, photoionization and photodissociation of gas species due to EUV, FUV, X-rays and cosmic rays, thermal energy exchange by gas-dust collisions, grain photoelectric heating of gas by FUV incident on PAHs and very small grains, and heating due to UV pumping of H$_2$ and exothermic chemical reactions. Cooling of gas is by line emission from ions, atoms and moleules. The chemical network is constructed to include the chemistry of the main gas cooling species, $\sim 120$ species made of H, He, C, N, O, Ne, S, Mg, Fe, Si and Ar with $\sim$ 800 chemical and photochemical reactions \citep{Gorti2011}. Grain surface chemistry is not treated explicitly, it is assumed that ices form at grain temperatures below the freeze-out temperatures of relevant species (e.g., CO, H$_2$O, CH$_4$) and include thermal desorption and photodesorption. 

Gas and dust are treated independently and the models allow for their different spatial distributions. A mixture of dust chemical compositions and a range of grain size distributions can be treated. Here we have assumed that all dust is comprised of silicates and the grain size distribution is determined from the SED fitting \citep{Lieman2016}. Gas opacities at each spatial location $(r,z)$ are calculated by dividing the FUV band into 9 bins including Lyman alpha, and absorption cross sections in each band are calculated by integrating the stellar spectrum with the photo-cross sections from the LAMDA\footnote{Leiden Atomic and Molecular Database, https://home.strw.leidenuniv.nl/~moldata/}
database for all available chemical species.  Dust radiative transfer is usually treated using a 1+1D construct, but is simple for the optically thin disks modeled here. Line radiative transfer (for calculating cooling in the thermochemical models) uses an escape probability formalism, explicitly computes the level populations of all coolants and is a full, non-LTE treatment. We note that the model results in this work are subsequently processed through the nLTE code LIME to generate emission maps for comparison with the ALMA data. 

The main model inputs are the stellar parameters (mass, spectrum, high energy flux), the dust disk parameters (constrained by the continuum emission and SED fitting) and the gas surface density distribution. Since the former are known or previously determined,  the surface density distribution, and hence mass, is the only variable in our modeling. Each profile thus generates a unique disk density, temperature and chemical density distribution as a function of spatial location $(r,z)$; this is then compared with the ALMA data to find the best fit model disk mass. 

\section{Drag force vs radiation pressure}

Assuming an infinite collisional cascade as in \citet{Wyatt2007} with
sizes ranging from the smallest sub-micron dust to km sizes, the resulting grain size distribution is such that the area/mass ratio is highest in the smallest grains. In a debris disk devoid of gas, the smallest grains are subject to removal by radiation pressure. In the 
presence of gas, however, the increased collisional coupling of small grains to gas molecules could result in their retention (in disks around A-stars, gravity begins to be become important for grains larger than $\sim 5\mu m$, e.g., \cite{wyatt2008}).  For gas drag  to be larger than radiation pressure, 
\begin{equation}
\frac{8\sqrt{2}\pi}{3} \rho_{gas} a^2 c_s \Delta v >
 \frac{L_* Q a^2}{4 \pi r^2 c}
\end{equation}
where the term on the left is the Epstein drag force and the right term is the force due to radiation pressure. Here, $\rho_{gas}, c_s$ are the gas density and thermal speed, $a$ is the grain size, $Q$ its absorption coefficient, $L_*$ the stellar luminosity, $r$ the distance to the star and $c$ the speed of light. Note that both forces depend on the grain area.  Assuming that the relative velocity between the gas and dust  $\Delta v \sim c_s$, and that $Q\sim 1$, the number density required for retaining collisionally generated dust is
\begin{equation}
n_{gas} (r) > 1.4\times 10^{10} \left(\frac{{\rm au}}{r}\right)^2 \left(\frac{20\ {\rm K}}{T_{gas}}\right) \left(\frac{L_*}{10 L_{\odot}}\right) \ {\rm \ cm}^{-3}
\end{equation}
While  $Q\sim1$  is valid for micron-sized grains at wavelengths where the stellar flux dominates, smaller sub-micron grains may have smaller values of $Q$ and may be retained at  densities lower than above. 
The density can be easily integrated to give required dust masses ($M_{disk} = \int 2\pi r dr \int dz\ \rho_{gas}$) for disks of $\sim 200$ au radii as in the present sample and by assuming a vertical extent $h/r\sim 0.1$; we therefore estimate that primordial disks with masses $M_{disk} \gtrsim 5$M$_{\oplus}$ of H$_2$ will begin to retain small dust grains as they are not removed by radiation pressure and remain collisionally coupled to the gas.  The dust generated by collisional cascades in such massive disks would accumulate until the disk becomes optically thick, and they would no longer resemble debris disks.

\end{document}